\documentclass[aps,pra,twocolumn,nobibnotes,superscriptaddress,nofootinbib,10pt]{revtex4-1} 
\usepackage[latin1]{inputenc}
\usepackage{amsmath,amssymb}
\usepackage{mathrsfs} 
\usepackage{graphicx,color,colortbl}
\usepackage{rotating,array,tabularx,booktabs}
\usepackage{varwidth,xcolor}
\usepackage{placeins}
\usepackage{siunitx}
\usepackage[normalem]{ulem}

\DeclareGraphicsExtensions{.pdf,.PDF,.png,.PNG}
\allowdisplaybreaks

\newcommand{\abs}[1]{\lvert#1\rvert}%
\newcommand{\ZT}[1]{\textquotedblleft#1\textquotedblright}%
\newcommand{\ww}{\vec{u}}%
\newcommand{\ws}{u}%

\begin{document}

\title{Propulsion of bullet- and cup-shaped nano- and microparticles by traveling ultrasound waves}

\author{Johannes Vo\ss{}}
\affiliation{Institut f\"ur Theoretische Physik, Center for Soft Nanoscience, Westf\"alische Wilhelms-Universit\"at M\"unster, 48149 M\"unster, Germany}

\author{Raphael Wittkowski}
\email[Corresponding author: ]{raphael.wittkowski@uni-muenster.de}
\affiliation{Institut f\"ur Theoretische Physik, Center for Soft Nanoscience, Westf\"alische Wilhelms-Universit\"at M\"unster, 48149 M\"unster, Germany}

\begin{abstract}
The propulsion of colloidal particles via planar traveling ultrasound waves has attracted increasing attention in recent years. A frequently studied type of particles is bullet-shaped and cup-shaped nano- and microparticles. Based on acoustofluidic simulations, this article investigates how the propulsion of bullet-shaped particles depends on their length and diameter, where cup-shaped particles are included as limiting cases corresponding to the smallest particle length. The structure of the flow field generated by the particles is discussed and it is shown that the particles' propulsion strength increases with their length and diameter. When varying the diameter, we observed also a sign change of the propulsion. This work complements previous experimental studies that have addressed such particles only for particular aspect ratios, and the provided understanding of how the propulsion of the particles depends on their dimensions will prospectively be helpful for the choice of particle shapes that are most suitable for future experimental studies.   
%
%
\end{abstract}
\maketitle

\section{Introduction}
Since their discovery in 2012 \cite{WangCHM2012}, ultrasound-propelled nano- and microparticles are a rapidly growing target for research \cite{WangCHM2012,GarciaGradillaEtAl2013,AhmedEtAl2013,NadalL2014,WuEtAl2014,WangLMAHM2014,GarciaGradillaSSKYWGW2014,BalkEtAl2014,AhmedGFM2014,AhmedLNLSMCH2015,WangDZSSM2015,EstebanFernandezdeAvilaMSLRCVMGZW2015,WuEtAl2015a,WuEtAl2015b,RaoLMZCW2015,Kiristi2015,KimGLZF2016,EstebanEtAl2016,SotoWGGGLKACW2016,AhmedWBGHM2016,KaynakONNLCH2016,UygunEtAl2017,EstebanFernandezEtAl2017,RenZMXHM2017,KaynakONLCH2017,CollisCS2017,ZhouZWW2017,ZhouYWDW2017,ChenEtAl2018,HansenEtAl2018,SabrinaTABdlCMB2018,AhmedBJPDN2016,Zhou2018,WangGWSGXH2018,EstebanEtAl2018,LuSZWPL2019,TangEtAl2019,QualliotineEtAl2019,GaoLWWXH2019,RenEtAl2019,BhuyanDBSGB2019,AghakhaniYWS2020,LiuR2020,VossW2020,ValdezLOESSWG2020,DumyJMBGMHA2020,McneillNBM2020,McneillSWOLNM2021,VossW2021,MohantyEtAl2021,LiMMOP2021,VossW2022orientation,VossW2022acoustica,VossW2022acousticb,VossW2022microspinner}.  
The research on this type of artificial motile particles is driven by the prospect of applying these particles in the future as useful nano- or microdevices that can perform medical tasks \cite{LiEFdAGZW2017,PengTW2017,SotoC2018,WangGZLH2020,WangZ2021,Leal2021,NitschkeW2021} or form active materials with intriguing properties \cite{JunH2010,McdermottKDKWSV2012,FratzlFKS2021}.    
So far, research on ultrasound-propelled nano- and microparticles has mainly been experimental \cite{WangCHM2012,GarciaGradillaEtAl2013,AhmedEtAl2013,WuEtAl2014,WangLMAHM2014,GarciaGradillaSSKYWGW2014,BalkEtAl2014,AhmedGFM2014,WangDZSSM2015,EstebanFernandezdeAvilaMSLRCVMGZW2015,WuEtAl2015a,WuEtAl2015b,LILXKLWW2015,AhmedLNLSMCH2015,EstebanEtAl2016,SotoWGGGLKACW2016,AhmedWBGHM2016,KaynakONNLCH2016,UygunEtAl2017,KaynakONLCH2017,EstebanFernandezEtAl2017,RenZMXHM2017,ZhouYWDW2017,HansenEtAl2018,SabrinaTABdlCMB2018,AhmedBJPDN2016,ZhouZWW2017,WangGWSGXH2018,EstebanEtAl2018,RenWM2018,Zhou2018,TangEtAl2019,QualliotineEtAl2019,GaoLWWXH2019,RenEtAl2019,AghakhaniYWS2020,LiuR2020,ValdezLOESSWG2020,DumyJMBGMHA2020}, but there have also been some publications that are based on acoustofluidic simulations \cite{KaynakONNLCH2016,AhmedBJPDN2016,Zhou2018,SabrinaTABdlCMB2018,WangGWSGXH2018,TangEtAl2019,RenEtAl2019,VossW2020,VossW2021,VossW2022orientation,VossW2022acoustica,VossW2022acousticb,VossW2022microspinner} and two studies that rely on analytical approaches \cite{NadalL2014,CollisCS2017}.  

Despite the intensive research on acoustically propelled particles, the understanding of these particles is still very limited. For example, little is known about how the propulsion of the particles depends on their shape. 
A reason for this limited knowledge is the fact that research on these particles is mainly experimental and it is much more difficult to vary the particle shape in experiments than in theory. 
Up to now, mostly bullet-shaped particles have been studied \cite{WangCHM2012,AhmedEtAl2013,GarciaGradillaEtAl2013,NadalL2014,GarciaGradillaSSKYWGW2014,WangLMAHM2014,AhmedGFM2014,BalkEtAl2014,WuEtAl2015a,EstebanFernandezdeAvilaMSLRCVMGZW2015,XuEtAl2015,Kiristi2015,EstebanEtAl2016,ZhouYWDW2017,ZhouZWW2017,EstebanFernandezEtAl2017,UygunEtAl2017,Zhou2018,HansenEtAl2018,WangGWSGXH2018,EstebanEtAl2018,BeltranEtAl2019,QualliotineEtAl2019,DumyJMBGMHA2020,McneillSWOLNM2021,AhmedWBGHM2016,CollisCS2017}. These are rod-shaped particles with a concave end and a convex end. 
Some studies have addressed hemi-spherical cups (nanoshells) \cite{SotoWGGGLKACW2016,TangEtAl2019,VossW2020}, which can be interpreted as the limit of a bullet-shaped particle with hemi-spherical ends for decreasing length of the cylindrical part between the two ends. 
There are also studies on cone-shaped particles \cite{VossW2020,VossW2021,VossW2022orientation,VossW2022acoustica,VossW2022acousticb}, a gear-shaped particle \cite{KaynakONNLCH2016,KaynakONLCH2017,SabrinaTABdlCMB2018}, and microspinners \cite{VossW2022microspinner}. 

Although bullet-shaped particles are frequently used, only particles with a few particular diameters and lengths have been realized \cite{WangCHM2012,AhmedEtAl2013,GarciaGradillaEtAl2013,NadalL2014,GarciaGradillaSSKYWGW2014,WangLMAHM2014,AhmedGFM2014,BalkEtAl2014,WuEtAl2015a,EstebanFernandezdeAvilaMSLRCVMGZW2015,XuEtAl2015,Kiristi2015,EstebanEtAl2016,ZhouYWDW2017,ZhouZWW2017,EstebanFernandezEtAl2017,UygunEtAl2017,Zhou2018,HansenEtAl2018,WangGWSGXH2018,EstebanEtAl2018,BeltranEtAl2019,QualliotineEtAl2019,DumyJMBGMHA2020,McneillSWOLNM2021,AhmedWBGHM2016,CollisCS2017}. 
Furthermore, the shape of the ends of the particles used in experiments was either hemi-spherical \cite{GarciaGradillaEtAl2013,AhmedEtAl2013,GarciaGradillaSSKYWGW2014,WangLMAHM2014,BalkEtAl2014,WuEtAl2015a,EstebanFernandezdeAvilaMSLRCVMGZW2015,Kiristi2015,EstebanEtAl2016,AhmedWBGHM2016,ZhouYWDW2017,EstebanFernandezEtAl2017,HansenEtAl2018,EstebanEtAl2018,QualliotineEtAl2019} or irregular due to a less-controlled fabrication process \cite{WangCHM2012,AhmedEtAl2013,GarciaGradillaEtAl2013,GarciaGradillaSSKYWGW2014,WangLMAHM2014,AhmedGFM2014,BalkEtAl2014,WuEtAl2015a,EstebanFernandezdeAvilaMSLRCVMGZW2015,XuEtAl2015,Kiristi2015,EstebanEtAl2016,ZhouYWDW2017,ZhouZWW2017,EstebanFernandezEtAl2017,UygunEtAl2017,Zhou2018,HansenEtAl2018,EstebanEtAl2018,BeltranEtAl2019,QualliotineEtAl2019,DumyJMBGMHA2020,McneillSWOLNM2021}. 
In the analytical study \cite{CollisCS2017}, bullet-shaped particles with hemi-spherical and hemi-spheroidal ends were studied, where the concavity and convexity of the ends were varied. 
This study found that the propulsion velocity depends strongly on the concavity and convexity of the ends. 
When varying the shape of the ends, even a change in the propulsion direction of the particle was observed. 
For the hemi-spherical-cup particles \cite{SotoWGGGLKACW2016,TangEtAl2019,VossW2020}, the opposite propulsion direction as for bullet-shaped particles with hemi-spherical ends was found. 
A problem additional to this limited insight into the shape-dependence of the acoustic propulsion is the fact that all except for a few \cite{AhmedBJPDN2016,VossW2021,VossW2022acoustica,VossW2022acousticb,VossW2022orientation,VossW2022microspinner} previous studies addressed particles in a standing ultrasound wave, whereas traveling ultrasound waves are much more relevant for the envisaged future applications of acoustically propelled particles \cite{VossW2020,VossW2021,VossW2022orientation}. 
  
In this article, we aim at extending the understanding of the effect the particle shape has on the particles' propulsion. 
We focus on bullet-shaped nano- and microparticles with hemi-spherical ends, as they have been applied in previous experiments \cite{GarciaGradillaEtAl2013,AhmedEtAl2013,GarciaGradillaSSKYWGW2014,WangLMAHM2014,BalkEtAl2014,WuEtAl2015a,EstebanFernandezdeAvilaMSLRCVMGZW2015,Kiristi2015,EstebanEtAl2016,AhmedWBGHM2016,ZhouYWDW2017,EstebanFernandezEtAl2017,HansenEtAl2018,EstebanEtAl2018,QualliotineEtAl2019}, and study, based on direct acoustofluidic simulations, how the flow field generated by the particles and the strength of their propulsion depend on the length and diameter of the particles. 
By including the limit of very short particles, where the concave and convex hemi-spherical ends get very close, we cover also hemi-spherical-cup particles. This allows us to understand at which particle length the sign change of the propulsion velocity, which is indicated by the previous experimental work, occurs. 
Different from most previous studies, we consider a traveling instead of a standing ultrasound wave to get more insight into this particularly application-relevant case.

\section{\label{methods}Methods}
For consistency and to rely on methods that have been proven to be successful, our methodology is adopted from Ref.\ \cite{VossW2020}. 

\subsection{Setup}
\begin{figure}[htb]
\centering
\includegraphics[width=\linewidth]{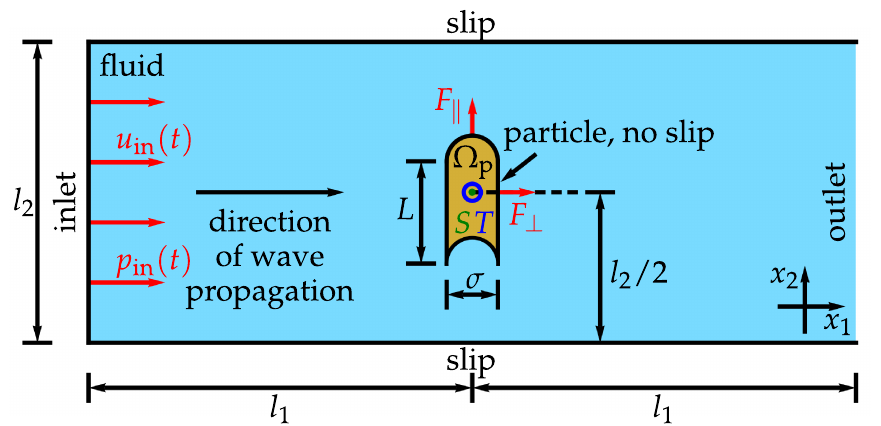}%
\caption{\label{fig:setup}Setup for the simulations.}
\end{figure}

An overview of the simulated system is given in Fig.\ \ref{fig:setup}. 
The system includes a rectangular fluid-filled domain and a bullet-shaped particle in the center of the domain.
We place the particle such that its center of mass $\mathrm{S}$ coincides with the geometric center of the rectangular domain. 
The particle's shape is fixed, i.e., we consider a solid particle.
It consists of a cylindrical part and two hemi-spherical ends of which one is concave and one is convex. 
This shape is motivated by previous experimental studies \cite{WangCHM2012,GarciaGradillaEtAl2013,AhmedEtAl2013,AhmedGFM2014,BalkEtAl2014,WangLMAHM2014,EstebanFernandezdeAvilaMSLRCVMGZW2015,AhmedWBGHM2016,ZhouZWW2017,WangGWSGXH2018,ZhouYWDW2017,Zhou2018,DumyJMBGMHA2020,SotoWGGGLKACW2016,TangEtAl2019,VossW2020}.
We orientate the particle in such a way that the vector from the center of the lower end of the particle to the center of its upper end points in the $x_2$-direction. 
The rectangular domain has width $2l_1$ (oriented parallel to the $x_1$-axis) and height $l_2=\SI{200}{\micro\metre}$ (oriented parallel to the $x_2$-axis) and the particle has variable diameter $\sigma\in [0.5,5]\,\SI{}{\micro\metre}$ and a cylindrical part of variable length $L\in [0.1,5]\,\SI{}{\micro\metre}$.

Because of its general relevance, we choose water as the fluid. When the simulation starts at time $t=0$, the water shall have a vanishing velocity field $\ww_0=\vec{0}\,\SI{}{\metre\,\second^{-1}}$. 
Moreover, the water shall initially be at standard temperature $T_0=\SI{293.15}{\kelvin}$ and standard pressure $p_0=\SI{101325}{\pascal}$, where it has mass density $\rho_0=\SI{998}{\kilogram\,\metre^{-3}}$, shear viscosity $\nu_\mathrm{s}=\SI{1.002}{\milli\pascal\,\second}$, bulk viscosity $\nu_\mathrm{b}=\SI{2.87}{\milli\pascal\,\second}$, and sound velocity $c_\mathrm{f}=\SI{1484}{\metre\,\second^{-1}}$. 

A planar traveling ultrasound wave enters the system at its left edge (inlet), propagates in the $x_1$-direction, interacts with the particle, and can leave the system at its right edge (outlet). 
At the inlet, we, therefore, prescribe a time-dependent velocity $\ws_{\mathrm{in}}(t)=\Delta u \sin(2\pi f t)$ and pressure $p_{\mathrm{in}}(t)=\Delta p \sin(2\pi f t)$, where $\Delta u=\Delta p / (\rho_0 c_{\mathrm{f}})$ is the velocity amplitude of the entering ultrasound wave, $\Delta p=\SI{10}{\kilo\pascal}$ its pressure amplitude, and $f=\SI{1}{\MHz}$ its frequency.
For the lower and upper edges of the simulation domain, we choose slip boundary conditions, and we use no-slip boundary conditions at the boundary of the particle domain $\Omega_\mathrm{p}$. 
Our choice for the pressure amplitude $\Delta p$ corresponds to an acoustic energy density $E=\Delta p^2/(2 \rho_0 c_{\mathrm{f}}^2)=\SI{22.7}{\milli\joule\,\metre^{-3}}$ that is considered to be harmless for humans and thus suitable for diagnostic applications \cite{BarnettEtAl2000}. 
The choice for the frequency $f$ is slightly lower than the frequencies that are common in sonography. 
For the wavelength of the ultrasound, we obtain $\lambda=c_\mathrm{f}/f=\SI{1484}{\milli\metre}$. 
As in Refs.\ \cite{VossW2020,VossW2021,VossW2022orientation,VossW2022acousticb,VossW2022microspinner}, we choose the width of the simulation domain so that $l_1=\lambda/4$. 

The interaction of the ultrasound with the particle causes forces and torques acting on the particle. 
We are interested in the stationary time-averaged propulsion force $F_\parallel$ that is oriented parallel to the particle, the stationary time-averaged propulsion force $F_\perp$ that is oriented perpendicular to the particle, and the stationary time-averaged propulsion torque $T$ that tends to rotate the particle in the $x_1$-$x_2$-plane. 
Both force components and the torque act on the center of mass $\mathrm{S}$ and are measured in the laboratory frame.

\subsection{Parameters}
Table \ref{tab:Parameters} shows the names and symbols of the parameters that are relevant for our simulations and their values that we have chosen in analogy to the values used in Ref.\ \cite{VossW2020}.
\begin{table*}[htb]
\caption{\label{tab:Parameters}Relevant parameters and the corresponding values.
For consistency, the values are chosen analogously to Ref.\ \cite{VossW2020}.}%
\centering
\begin{ruledtabular}
\begin{tabular}{llll}%
\textbf{Name} & \textbf{Symbol} & \textbf{Value} & \textbf{Remark}\\
\hline
Particle diameter & $\sigma$ & $0.5$-$\SI{5}{\micro\metre}$ & \\
Particle length & $L$ & $0.1$-$\SI{5}{\micro\metre}$ & \\
Sound frequency & $f$ & $\SI{1}{\MHz}$ & \\
Speed of sound & $c_\mathrm{f}$ & $\SI{1484}{\metre\,\second^{-1}}$ & Corresponds to $T_0$ and $p_0$ \\
Time period of sound & $\tau=1/f$ & $\SI{1}{\micro\second}$ & \\
Wavelength of sound & $\lambda=c_\mathrm{f}/f$ & $\SI{1.484}{\milli\metre}$ & \\
Temperature of fluid & $T_0$ & $\SI{293.15}{\kelvin}$ & \\
Mean mass density of fluid & $\rho_0$ & $\SI{998}{\kilogram\,\metre^{-3}}$ & Corresponds to $T_0$ and $p_0$ \\
Mean pressure of fluid & $p_{0}$ & $\SI{101325}{\pascal}$ & \\
Initial velocity of fluid & $\ww_{0}$ & $\vec{0}\,\SI{}{\metre\,\second^{-1}}$ & \\
Sound pressure amplitude & $\Delta p$ & $\SI{10}{\kilo\pascal}$ & \\
Flow velocity amplitude & $\Delta u=\Delta p / (\rho_0 c_\mathrm{f})$ & $\SI{6.75}{\milli\metre\,\second^{-1}}$ & \\
Acoustic energy density & $E=\Delta p^2/(2 \rho_0 c_{\mathrm{f}}^2)$ & $\SI{22.7}{\milli\joule\,m^{-3}}$ & \\
Shear/dynamic viscosity of fluid & $\nu_{\mathrm{s}}$ & $\SI{1.002}{\milli\pascal\,\second}$ & Corresponds to $T_0$ and $p_0$ \\
Bulk/volume viscosity of fluid & $\nu_{\mathrm{b}}$ & $\SI{2.87}{\milli\pascal\,\second}$ & Interpolated from Tab.\ 1 in Ref.\ \cite{HolmesPP2011} for $T_0$ and $p_0$ \\
Inlet-particle distance & $l_{1}$ & $\lambda/4$ & \\
Domain width & $l_2$ & $\SI{200}{\micro\metre}$ & \\
Mesh-cell size & $\Delta x$ & $\SI{15}{\nano \metre}$-$\SI{1}{\micro \metre}$ & \\
Time-step size & $\Delta t$ & $1$-$\SI{10}{\pico \second}$ & \\
Simulation duration & $t_{\mathrm{max}}$ & $\geqslant 500\tau$ & \\
Euler number & $\mathrm{Eu}$ &  $\SI{2.2}{\cdot10^5}$ & \\
Helmholtz number & $\mathrm{He}$ & $\SI{3.37}{\cdot10^{-4}}$-$\SI{3.37}{\cdot10^{-3}}$ & \\
Bulk Reynolds number &  $\mathrm{Re}_\mathrm{b}$ & $\SI{1.17}{\cdot10^{-3}}$-$\SI{1.17}{\cdot10^{-2}}$ & \\
Shear Reynolds number &  $\mathrm{Re}_\mathrm{s}$ & $\SI{3.36}{\cdot10^{-3}}$-$\SI{3.36}{\cdot10^{-2}}$ & \\
Particle Reynolds number &  $\mathrm{Re}_\mathrm{p}$ & $<\SI{2}{\cdot10^{-6}}$ & 
\end{tabular}%
\end{ruledtabular}%
\end{table*}

\subsection{Acoustofluidic simulations}
We use direct computational fluid dynamics simulations to simulate the propagation of the ultrasound and its interaction with the particle. 
They consist of solving the standard equations of fluid dynamics (continuity equation, compressible Navier-Stokes equations, linear constitutive equation for the pressure). 

The dimensionless numbers that result from these equations by nondimensionalization are the 
Euler number
\begin{equation}
\mathrm{Eu}=\frac{\Delta p}{\rho_0 \Delta u^2}\approx \SI{2.2}{\cdot10^{5}},
\end{equation}
Helmholtz number
\begin{equation}
\mathrm{He}=\frac{f \max\{L,\sigma\}}{c_\mathrm{f}}\approx \SI{3.37}{\cdot 10^{-4}}\text{-}\SI{3.37}{\cdot 10^{-3}},
\end{equation}
bulk Reynolds number
\begin{equation}
\mathrm{Re}_\mathrm{b}=\frac{\rho_0 \Delta u \max\{L,\sigma\}}{\nu_\mathrm{b}}\approx \SI{1.17}{\cdot10^{-3}}\text{-}\SI{1.17}{\cdot10^{-2}},
\end{equation}
and shear Reynolds number
\begin{equation}
\mathrm{Re}_\mathrm{s}=\frac{\rho_0 \Delta u \max\{L,\sigma\}}{\nu_\mathrm{s}}\approx \SI{3.36}{\cdot10^{-3}}\text{-}\SI{3.36}{\cdot10^{-2}}.
\end{equation}
Considering the motion of the particle in the fluid that can result from the propulsion force and torque acting on the particle, one can also define a particle Reynolds number
\begin{align}
\mathrm{Re}_\mathrm{p}&=\frac{\rho_0 }{\nu_\mathrm{s}} \max\{L,\sigma\}\sqrt{v_\parallel^2+v_\perp^2} <2\cdot 10^{-6},
\end{align}
where $v_\parallel$ and $v_\perp$ are the propulsion velocities that correspond to $F_\parallel$ and $F_\perp$, respectively (see Section \ref{A:velocities}).
See Ref.\ \cite{VossW2021} for an interpretation of these dimensionless numbers. 

We solve the equations of fluid dynamics with the finite volume software package OpenFOAM \cite{WellerTJF1998}. 
For spatial discretization of the studied system, we use a structured, mixed rectangular-triangular mesh with about 300,000 cells. The cell size $\Delta x$ is very small close to the particle and increases with the distance from the particle. 
For temporal discretization of the time range from $t=0$, where a simulation starts, to $t=t_\mathrm{max} \geqslant 500\tau$ with the period $\tau=1/f$ of the ultrasound wave, where the simulation ends, we use an adaptive time-step method.
The time-step size $\Delta t$ always has to fulfill the Courant-Friedrichs-Lewy condition
\begin{align}
C = c_\mathrm{f} \frac{\Delta t}{\Delta x} < 1 .
\end{align}
Since these simulations require a fine discretization in space and time, the typical computational expense of an individual simulation run amounts to $36,000$ CPU core hours.

\subsection{Propulsion force and torque}
From the results of the acoustofluidic simulations, we obtain the stationary time-averaged propulsion force components $F_\parallel$ and $F_\perp$ and propulsion torque $T$ by suitable integration of the stress tensor $\Sigma$ over the particle surface, locally averaging over one period, and extrapolating towards $t \to \infty$ (see Ref.\ \cite{VossW2020} for details on this procedure).  
By distinguishing the pressure component $\Sigma^{(p)}$ and the viscous component $\Sigma^{(v)}$ of the stress tensor $\Sigma=\Sigma^{(p)}+\Sigma^{(v)}$, we can split $F_\parallel$, $F_\perp$, and $T$ as $F_\parallel=F_{\parallel,p}+F_{\parallel,v}$, $F_\perp = F_{\perp,p}+F_{\perp,v}$, and $T=T_p + T_v$ into a pressure contribution (subscript \ZT{$p$}) that originates from $\Sigma^{(p)}$ and a viscous contribution (subscript \ZT{$v$}) that originates from $\Sigma^{(v)}$.

\subsection{\label{A:velocities}Translational and angular propulsion velocity}
Using the Stokes law \cite{HappelB1991}, we calculate from $F_\parallel$, $F_\perp$, and $T$ the corresponding translational propulsion velocity parallel to the particle's orientation $v_{\parallel}$, translational propulsion velocity perpendicular to the particle's orientation $v_{\perp}$, and angular propulsion velocity $\omega$.     
This transformation (see Ref.\ \cite{VossW2020} for details) involves the hydrodynamic resistance matrix $\boldsymbol{\mathrm{H}}$, which depends on the size and shape of the particle. 
For better reproducibility of our results, we here present the hydrodynamic resistance matrix for the particle studied in the present work, which can be calculated, e.g., with the software \texttt{HydResMat} \cite{VossW2018,VossJW2019}. 
When choosing the center of mass $\mathrm{S}$ as the reference point for the calculation of $\boldsymbol{\mathrm{H}}$ and assigning a thickness of $\sigma$ in the third dimension to the particle (see Ref.\ \cite{VossW2020} for details), $\boldsymbol{\mathrm{H}}$ has the general structure
\begin{align}
\boldsymbol{\mathrm{H}} = \begin{pmatrix}
\mathrm{K_{11}} & 0 & 0 & 0 & 0 & \mathrm{C_{31}}\\
 0 & \mathrm{K_{22}} & 0 & 0 & 0 & 0 \\
 0 & 0 & \mathrm{K_{33}} & \mathrm{C_{13}} & 0 & 0\\
 0 & 0 & \mathrm{C_{13}} & \mathrm{\Omega_{11}} & 0 & 0 \\
 0 & 0 & 0 & 0 & \mathrm{\Omega_{22}} & 0 \\
 \mathrm{C_{31}} & 0 & 0 & 0 & 0 &\mathrm{\Omega_{33}}
\end{pmatrix}.
\end{align}
The values of the nonzero elements are given in Tab.\ \ref{tab:resistance} for each particle diameter $\sigma$ and length $L$ considered in this work.
\begin{table*}[tb]
\centering
\caption{\label{tab:resistance}Nonzero elements of the hydrodynamic resistance matrix $\boldsymbol{\mathrm{H}}$ of the particle that is considered in this article (see Fig.\ \ref{fig:setup}).}
\begin{ruledtabular}
\begin{tabular}{cccccccccc}
$\boldsymbol{\sigma} / \SI{}{\textbf{\micro\metre}}$ & $\boldsymbol{L} / \SI{}{\textbf{\micro\metre}}$ & $\mathbf{K_{11}} / \SI{}{\textbf{\micro\metre}}$ & $\mathbf{K_{22}}/ \SI{}{\textbf{\micro\metre}}$ & $\mathbf{K_{33}}/ \SI{}{\textbf{\micro\metre}}$ & $\mathbf{C_{13}}/ \SI{}{\textbf{\micro\metre}^{\mathbf{2}}}$ & $\mathbf{C_{31}}/ \SI{}{\textbf{\micro\metre}^{\mathbf{2}}}$& $\mathbf{\Omega_{11}}/ \SI{}{\textbf{\micro\metre}^{\mathbf{3}}}$ & $\mathbf{\Omega_{22}}/ \SI{}{\textbf{\micro\metre}^{\mathbf{3}}}$ & $\mathbf{\Omega_{33}}/ \SI{}{\textbf{\micro\metre}^{\mathbf{3}}}$ \\
\hline
$\SI{0.5}{}$ & $\SI{5}{}$ & $\SI{21.91}{}$ & $\SI{15.36}{}$ & $\SI{21.68}{}$ & $\SI{-0.73}{}$ & $\SI{1.74}{}$ & $\SI{86.08}{}$ & $\SI{6.48}{}$ & $\SI{89.32}{}$ \\
$\SI{0.5}{}$ & $\SI{1}{}$ & $\SI{9.08}{}$ & $\SI{7.68}{}$ & $\SI{8.81}{}$ & $\SI{-0.42}{}$ & $\SI{0.84}{}$ & $\SI{3.17}{}$ & $\SI{1.74}{}$ & $\SI{3.54}{}$\\
$\SI{0.5}{}$ & $\SI{0.5}{}$ &$\SI{6.96}{}$ & $\SI{6.41}{}$ & $\SI{6.65}{}$ & $\SI{-0.31}{}$ & $\SI{0.66}{}$ & $\SI{1.23}{}$ & $\SI{1.13}{}$ & $\SI{1.42}{}$\\
$\SI{0.5}{}$ & $\SI{0.1}{}$ &$\SI{4.95}{}$ & $\SI{5.12}{}$ & $\SI{4.50}{}$ & $\SI{-0.20}{}$ & $\SI{0.51}{}$ & $\SI{0.44}{}$ & $\SI{0.61}{}$ & $\SI{0.53}{}$\\
$\SI{1}{}$ & $\SI{0.1}{}$ & $\SI{9.39}{}$ & $\SI{9.96}{}$ & $\SI{8.41}{}$ & $\SI{-0.76}{}$ & $\SI{2.01}{}$ & $\SI{3.09}{}$ & $\SI{4.35}{}$ & $\SI{3.72}{}$\\
$\SI{2}{}$ & $\SI{0.1}{}$ & $\SI{17.92}{}$ & $\SI{19.16}{}$ & $\SI{15.82}{}$ & $\SI{-2.58}{}$ & $\SI{7.68}{}$ & $\SI{22.37}{}$ & $\SI{31.81}{}$ & $\SI{27.03}{}$\\
$\SI{5}{}$ & $\SI{0.1}{}$ & $\SI{43.79}{}$ & $\SI{47.42}{}$ & $\SI{38.56}{}$ & $\SI{-15.29}{}$ & $\SI{47.95}{}$ & $\SI{333.19}{}$ & $\SI{472.74}{}$ & $\SI{405.73}{}$
\end{tabular} 
\end{ruledtabular}%
\end{table*}

\section{\label{results}Results and discussion}
In this section, we present our simulation results for the time-averaged stationary flow field that forms around a bullet-shaped particle as depicted in Fig.\ \ref{fig:setup} and for the strength of the associated time-averaged stationary propulsion of the particle. 
We varied the length $L\in[0.1,5]\,\SI{}{\micro\metre}$ of the particle while keeping its diameter constant at $\sigma=\SI{0.5}{\micro\metre}$, and we varied the particle's diameter $\sigma\in [0.5,5]\,\SI{}{\micro\metre}$ while keeping its length constant at $L=\SI{0.1}{\micro\metre}$, to study the influence of these parameters on the flow field and propulsion strength. 
In the first case, the particle is an increasingly elongated bullet, whereas in the second case, it is a hemi-spherical cup with increasing size.

\subsection{Flow field}
\begin{figure*}[tb]
\centering
\includegraphics[width=\linewidth]{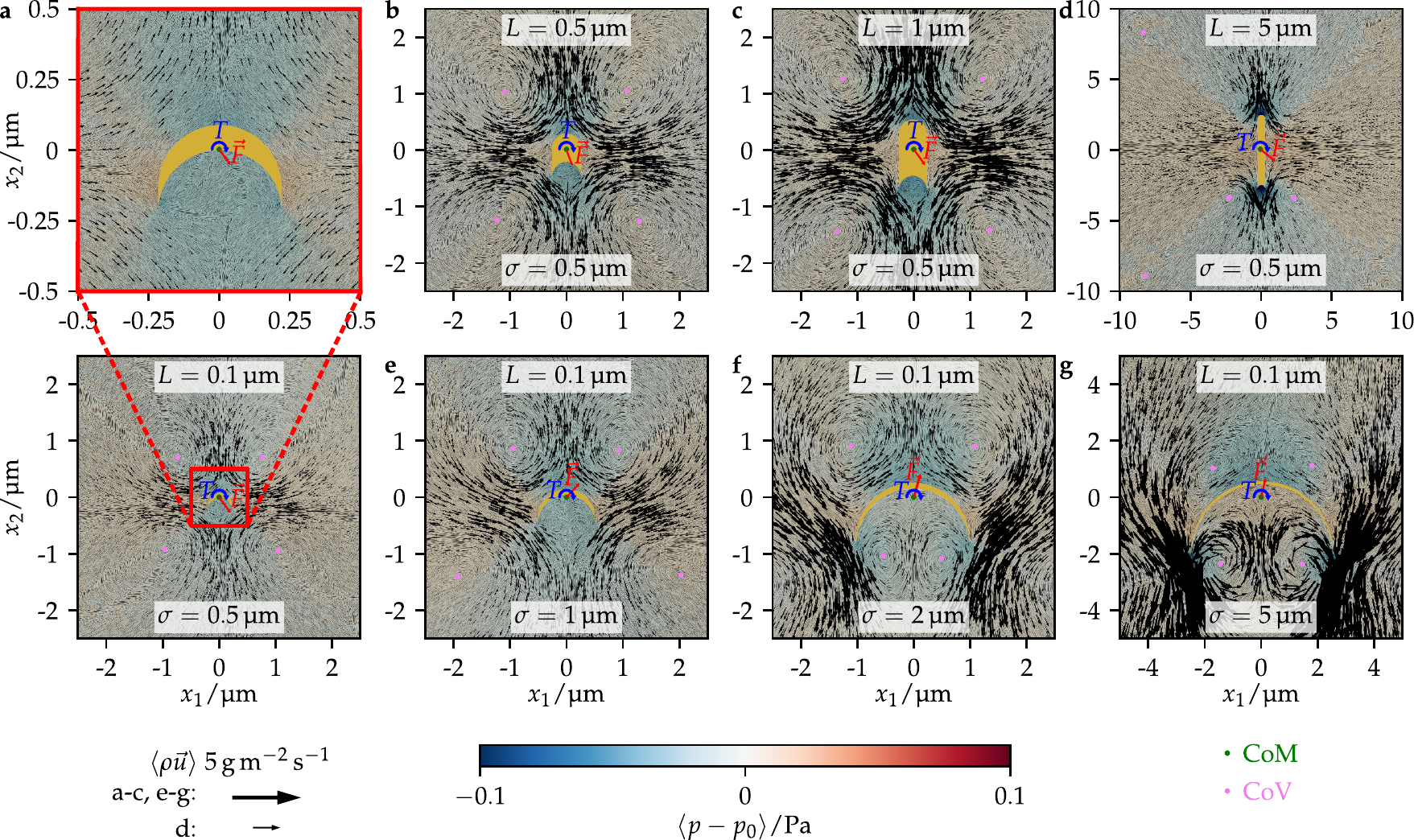}%
\caption{\label{fig:fig2}Time-averaged stationary mass-current density $\langle\rho\ww\rangle$ and reduced pressure $\langle p-p_{0}\rangle$ around bullet- and cup-shaped acoustically propelled particles (see Fig.\ \ref{fig:setup}) with different diameters $\sigma$ and lengths $L$. 
The particle's center of mass (CoM), the fluid's centers of vortices (CoV), and the orientations of the propulsion force $\vec{F}$ and torque $T$ that are exerted on the particle are indicated.} 
\end{figure*}

Our simulation results for the time-averaged stationary flow field that forms around the particle are shown in Fig.\ \ref{fig:fig2}.

For our discussion of the flow field, we consider a cup-shaped particle with diameter $\sigma=\SI{0.5}{\micro\metre}$ and length $L=\SI{0.1}{\micro \metre}$ (see Fig.\ \ref{fig:fig2}a) as a reference particle. 
We can see that the flow field around this particle is dominated by four large vortices, whose centers are located at the top left, top right, bottom left, and bottom right of the particle. Their distances from the center of mass of the particle are similar. 
In addition, there are two small and weak vortices near the concave end of the particle. 
The rotation directions of the large vortices are such that the fluid flows laterally towards the particle and above and below the particle away from it. As a consequence, the pressure is laterally increased and above and below the particle decreased.
Flow fields with a qualitatively similar structure (but for convex particle shapes without the small vortices) have already been observed for solid or hollowed-out half-sphere-shaped particles \cite{VossW2020} and for solid \cite{VossW2021,VossW2022orientation,VossW2022acoustica,VossW2022acousticb} or hollowed-out \cite{VossW2020} cone-shaped or triangular particles. 
In the present case, the centers of the large vortices above the particle are slightly nearer to the particle's center of mass than those below the particle.

\subsubsection{\label{sec:flowfieldL}Variation of length $L$}
First, we discuss the dependence of the particle's flow field on the particle's length $L$. 
When, starting at the reference particle, we increase $L$ from $L=\SI{0.1}{\micro\metre}$ to $L=\SI{1}{\micro\metre}$, the centers of the large vortices move slowly away from the center of mass of the particle, while the distances of the centers of the large vortices from the center of mass of the particle remain similar.
However, if we increase the length to $L=\SI{5}{\micro\metre}$, we obtain a qualitatively different flow field, which is less symmetric. 
While the centers of the four large vortices move further away from the center of mass of the particle, the two right vortex centers now move away much faster than the two left ones (in Fig.\ \ref{fig:fig2}d, the large right vortices are therefore outside of the plotted region). 
The distances of the centers of the large vortices from the center of mass of the particle are $\approx\SI{11.6}{\micro\metre}$ for the left vortices and $\approx\SI{33.2}{\micro\metre}$ for the right ones. 
This leads to an asymmetric arrangement of the vortices. 
Furthermore, two additional smaller vortices close to the particle occur now at its bottom left and bottom right. 
The distance of the two smaller vortices from the center of mass of the particle is $\approx\SI{4.1}{\micro\metre}$.
We can also see that, when increasing $L$ from $L=\SI{0.1}{\micro\metre}$ to $L=\SI{5}{\micro\metre}$, the flow near the particle becomes faster.

\subsubsection{\label{sec:flowfieldsigma}Variation of diameter $\sigma$}
Next, we discuss how the particle's flow field depends on the particle's diameter $\sigma$. 
When we increase $\sigma$ from $\sigma=\SI{0.5}{\micro\metre}$ to $\sigma=\SI{1}{\micro\metre}$, the centers of the large vortices again move slowly away from the center of mass of the particle. 
However, now the distance of the two lower vortex centers from the center of mass of the particle increases significantly faster than the distance of the two upper vortex centers.  
Increasing the diameter further to $\sigma=\SI{2}{\micro\metre}$ leads to a qualitatively different flow field. 
Now, the two large vortices below the particle vanish and the two small vortices inside the cavity of the particle increase and move further below the particle. 
Far below the particle, the fluid flows still away from the particle, but close to the particle the lower two vortices now lead to a flow that runs parallel to the symmetry axis of the particle towards the center of the particle, from there along the lower particle surface to the edge of the particle, then around the vortices below the particle, and again towards the center of the particle.
The distances of the centers of the four vortices from the center of mass of the particle are similar, but the vortex centers below the particle are nearer to each other than the upper vortex centers are. 
Further increasing the diameter to $\sigma=\SI{5}{\micro\metre}$ leads to larger distances of all vortex centers from the center of mass of the particle, where the vortex centers below the particle increase their distance from the center of mass of the particle faster than the vortex centers above the particle.
Thereby, the distances between the two upper and the two lower vortex centers become more similar. 
The qualitative structure of the flow field thus changes differently for increasing $\sigma$ than for increasing $L$. 
Nevertheless, also for increasing $\sigma$ from $\sigma=\SI{0.5}{\micro\metre}$ to $\sigma=\SI{5}{\micro\metre}$, we see that the flow near the particle becomes faster.

\subsection{Strength of propulsion}
\begin{figure*}[tb]
\centering
\includegraphics[width=\linewidth]{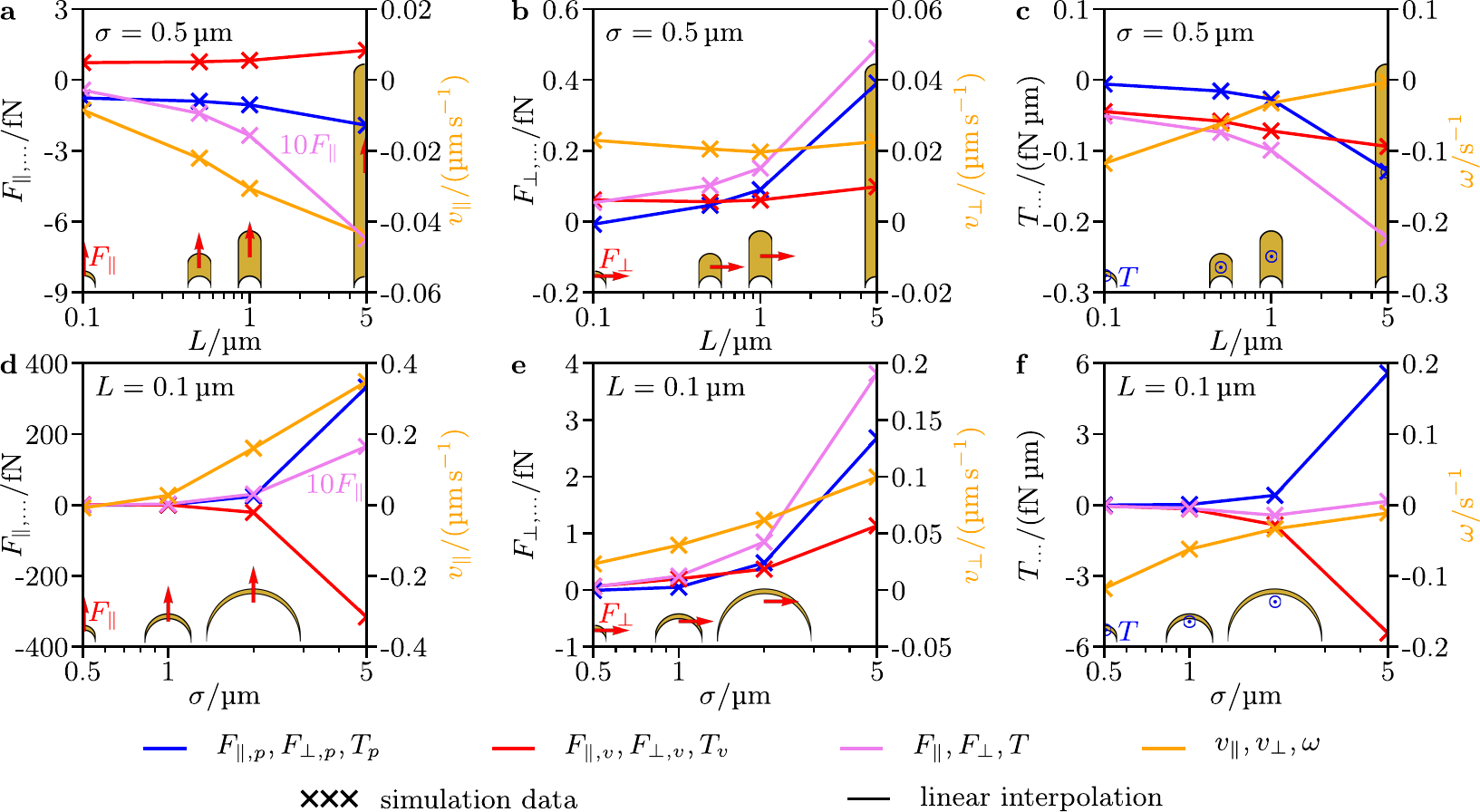}%
\caption{\label{fig:fig3}Simulation data for the time averaged propulsion forces $F_{\parallel}$ and $F_{\perp}$ and torque $T$ acting on the particle, their pressure components $F_{\parallel,p}$, $F_{\perp, p}$, and $T_p$, their viscous components $F_{\parallel, v}$, $F_{\perp, v}$, and $T_v$, and the corresponding translational propulsion velocities $v_\parallel$ and $v_\perp$ and angular propulsion velocity $\omega$ for various \textbf{a}-\textbf{c} particle lengths $L$ with $\sigma=\SI{0.5}{\micro\metre}$ and \textbf{d}-\textbf{f} particle diameters $\sigma$ with $L=\SI{0.1}{\micro\metre}$. 
In \textbf{a} and \textbf{d}, $10F_\parallel$ is shown instead of $F_\parallel$ for better visibility of the course of the function.} 
\end{figure*}

Figure \ref{fig:fig3} shows our simulation results for the time-averaged stationary propulsion forces $F_{\parallel}$ and $F_{\perp}$ and torque $T$ that act on the particle, their pressure components $F_{\parallel,p}$, $F_{\perp, p}$, and $T_p$, their viscous components $F_{\parallel, v}$, $F_{\perp, v}$, and $T_v$, and the corresponding translational propulsion velocities $v_\parallel$ and $v_\perp$ and angular propulsion velocity $\omega$ of the particle.

\subsubsection{\label{sec:propulsionL}Variation of length ${L}$}
Again, we first consider a variation of the particle's length $L$ while keeping the diameter $\sigma$ constant. 
 
To start, we focus on the parallel components of the propulsion (see Fig.\ \ref{fig:fig3}a). 
When $L$ increases, $F_{\parallel,p}$ decreases from $F_{\parallel,p}=\SI{-0.77}{\femto\newton}$ to $F_{\parallel,p}=\SI{-1.92}{\femto\newton}$. 
On the other hand, $F_{\parallel, v}$ increases from $F_{\parallel,v}=\SI{0.73}{\femto\newton}$ to $F_{\parallel,v}=\SI{1.25}{\femto\newton}$. 
Since $F_{\parallel,p}$ is dominant, the parallel propulsion force $F_{\parallel}$ decreases from $F_\parallel=\SI{-0.044}{\femto\newton}$ to $F_\parallel=\SI{-0.67}{\femto\newton}$. 
The parallel propulsion velocity $v_\parallel$ has a similar trend. It decreases from $v_\parallel=\SI{-0.009}{\micro\metre\,\second^{-1}}$ to $v_\parallel=\SI{-0.044}{\micro\metre\,\second^{-1}}$. 
The finding that the particle speed $\abs{v_\parallel}$ increases with $L$ is in agreement with the observation of Section \ref{sec:flowfieldL} that the flow near the particle becomes faster for increasing $L$.
Since $v_\parallel$ is always negative, the particle will move backward, i.e., towards its concave end. 
This finding is in line with available experimental results \cite{AhmedWBGHM2016}.
The same study, on the other hand, found a decrease in the particle speed for increasing $L$.
This differs from our results but could be explained by the differences in their setup compared to our study.
For example, their particle has a different diameter ($\sigma=\SI{0.3}{\micro\metre}$) than our particle, they used a standing instead of a traveling ultrasound wave, and their ultrasound had a different frequency ($f=\SI{3.77}{\MHz}$). 
 
Now, we consider the perpendicular components of the propulsion (see Fig.\ \ref{fig:fig3}b).
When $L$ is increased, the force $F_{\perp, p}$ increases from $F_{\perp,p}=\SI{-0.0076}{\femto\newton}$ to $F_{\perp,p}=\SI{0.39}{\femto\newton}$.
In contrast, $F_{\perp, v}$ only increases from $F_{\perp,v}=\SI{0.061}{\femto\newton}$ to $F_{\perp,v}=\SI{0.098}{\femto\newton}$. 
The function for the perpendicular propulsion force $F_{\perp}$ is therefore rather parallel to the curve for $F_{\perp, p}$ and increases from $F_\perp=\SI{0.053}{\femto\newton}$ to $F_\perp=\SI{0.49}{\femto\newton}$. 
Since the translational hydrodynamic resistance of the particle also increases with $L$, the perpendicular propulsion velocity $v_\perp$ is roughly constant at $v_\perp=\SI{0.02}{\micro\metre\,\second^{-1}}$.
 
Next, we address the rotational components of the propulsion (see Fig.\ \ref{fig:fig3}c). 
For increasing $L$, the components $T_p$ and $T_v$ as well as the propulsion torque $T$ are negative and further decreasing. $T_p$ decreases from $T_p=\SI{-0.006}{\femto\newton\,\micro\metre}$ to $T_p=\SI{-0.13}{\femto\newton\,\micro\metre}$, $T_v$ decreases from $T_v=\SI{-0.045}{\femto\newton\micro\metre}$ to $T_v=\SI{-0.094}{\femto\newton\,\micro\metre}$, and thus $T$ decreases from $T = \SI{-0.051}{\femto\newton\,\micro\metre}$ to $T=\SI{-0.23}{\femto\newton\,\micro\metre}$.   
The angular propulsion velocity $\omega$, however, increases from $\omega=\SI{-0.12}{\second^{-1}}$ to zero, since the angular hydrodynamic resistance $\Omega_{33}$ of the particle increases steeply with $L$ (see Tab.\ \ref{tab:resistance}). 
We can compare the, according to the amount, largest angular velocity $\omega=\SI{-0.12}{\second^{-1}}$ with the particle's rotation through Brownian motion. 
The latter is characterized by the particle's rotational diffusion coefficient $D_\mathrm{R}=(k_\mathrm{B}T_0/\nu_\mathrm{s})(\boldsymbol{\mathrm{H}}^{-1})_{66}=\SI{8.47}{\second^{-1}}$, corresponding to rotation in the $x_1$-$x_2$ plane, where $k_\mathrm{B}$ is the Boltzmann constant. 
Brownian rotation with this rotational diffusion coefficient is associated with a reorientation of the particle on the timescale $\SI{0.12}{\second}$.
On the other hand, the angular velocity $\omega=\SI{-0.12}{\second^{-1}}$ corresponds to a reorientation by 90\textdegree{} in about $\SI{13}{\second}$. 
This shows that the Brownian rotation is dominant and the rotational acoustic propulsion can be ignored here, which is in line with the findings of previous studies that addressed other kinds of ultrasound-propelled particles \cite{VossW2020,VossW2021,VossW2022orientation,VossW2022acoustica,VossW2022acousticb}.

For the limiting case of a hemi-spherical cup ($L=\SI{0.1}{\micro\metre}$), we thus find that, according to the amount, the parallel propulsion velocity is particularly small, the perpendicular propulsion velocity is similar to that of longer bullet-shaped particles, and the angular propulsion is particularly large. 
We thus conclude that hemi-spherical cups are less suitable than longer bullet-shaped particles for transporting drugs or other cargo to a target, as it is a future scenario of nanomedicine \cite{GarciaGradillaEtAl2013,GarciaGradillaSSKYWGW2014,Kiristi2015,EstebanEtAl2016,ZhouYWDW2017,EstebanFernandezEtAl2017,UygunEtAl2017,HansenEtAl2018}, since the former particles exhibit slower translational but faster rotational motion.

\subsubsection{\label{sec:propulsionsigma}Variation of diameter ${\sigma}$}
Again, we proceed to a variation of the particle's diameter $\sigma$ while keeping the length $L$ constant.
 
At first, we consider the parallel components of the propulsion (see Fig.\ \ref{fig:fig3}d). 
When $\sigma$ increases, $F_{\parallel,p}$ increases from $F_{\parallel,p} = \SI{-0.77}{\femto\newton}$ to $F_{\parallel,p} = \SI{332.67}{\femto\newton}$. 
On the contrary, $F_{\parallel, v}$ decreases from $F_{\parallel,v} = \SI{0.73}{\femto\newton}$ to $F_{\parallel,v} = \SI{-316.15}{\femto\newton}$.
Since $F_{\parallel,p}$ is again dominant, the parallel propulsion force $F_{\parallel}$ increases from $F_\parallel=\SI{-0.044}{\femto\newton}$ to $F_\parallel=\SI{16.52}{\femto\newton}$.
The parallel propulsion velocity $v_\parallel$ also increases, starting with $v_\parallel=\SI{-0.0085}{\micro\metre\,\second^{-1}}$ and ending with $v_\parallel=\SI{0.35}{\micro\metre\,\second^{-1}}$.
This overall increase of the particle speed with $\sigma$ is in line with the observation of Section \ref{sec:flowfieldsigma} that the flow near the particle becomes faster for increasing $\sigma$.
While the particle moves backward -- like the more elongated bullet-shaped particles that are considered in Section \ref{sec:propulsionL} -- for diameter $\sigma=\SI{0.5}{\micro\metre}$, it starts to move forward and rapidly increases its forward speed when $\sigma$ is increased. This increase of $v_\parallel$ is even steeper than the increase of $v_\parallel$ with $L$ (see Section \ref{sec:propulsionL}).  
The sign change of the propulsion parallel to the particle orientation must occur somewhere between $\sigma=\SI{0.5}{\micro\metre}$ and $\sigma=\SI{1}{\micro\metre}$.
These findings are consistent with the experimental results reported in Ref.\ \cite{SotoWGGGLKACW2016}, where a hemi-spherical-cup-shaped particle and a bullet-shaped particle were found to propel in opposite directions. 
The different signs of the propulsion were theoretically explained in Ref.\ \cite{CollisCS2017}. 
Using an analytical approach, this study shows that such a sign change can occur at $\beta \in \mathcal{O}(1)$ with the acoustic Reynolds number $\beta=\pi\rho_0\sigma^2 f/(2\nu_\mathrm{s})$. 
The occurrence of the sign change between $\sigma=\SI{0.5}{\micro\metre}$ and $\sigma=\SI{1}{\micro\metre}$ in our simulations corresponds to a sign change between $\beta=\SI{0.4}{}$ and $\beta=\SI{1.6}{}$ in terms of the acoustic Reynolds number, which is consistent with the condition $\beta \in \mathcal{O}(1)$ presented in Ref.\ \cite{CollisCS2017}. 
 
Now, we consider the perpendicular components of the propulsion (see Fig.\ \ref{fig:fig3}e).
When $\sigma$ is increased, all perpendicular quantities $F_{\perp, p}$, $F_{\perp, v}$, $F_{\perp}$, and $v_\perp$ grow. 
$F_{\perp, p}$ increases from $F_{\perp,p}=\SI{-0.0076}{\femto\newton}$ to $F_{\perp,p}=\SI{2.68}{\femto\newton}$, $F_{\perp, v}$ increases from $F_{\perp,v}=\SI{0.061}{\femto\newton}$ to $F_{\perp,v}=\SI{1.13}{\femto\newton}$, the perpendicular propulsion force $F_{\perp}$ increases from $F_\perp=\SI{0.053}{\femto\newton}$ to $F_\perp=\SI{3.82}{\femto\newton}$, and the perpendicular propulsion velocity $v_\perp$ increases from $v_\perp = \SI{0.023}{\micro\metre\,\second^{-1}}$ to $v_\perp=\SI{0.10}{\micro\metre\,\second^{-1}}$.

Finally, we address the rotational components of the propulsion (see Fig.\ \ref{fig:fig3}f). 
For increasing $\sigma$, the torque $T_p$ increases from $T_p=\SI{-0.006}{\femto\newton\,\micro\metre}$ to $T_p=\SI{5.58}{\femto\newton\,\micro\metre}$, whereas $T_v$ decreases from $T_v=\SI{-0.045}{\femto\newton\,\micro\metre}$ to $T_v=\SI{-5.43}{\femto\newton\,\micro\metre}$. 
As a consequence, the propulsion torque $T$ is rather constant and close to zero. 
However, the angular propulsion velocity $\omega$ strongly increases from $\omega=\SI{-0.12}{\second^{-1}}$ to $\omega=\SI{-0.011}{\second^{-1}}$, since the angular hydrodynamic resistance of the particle increases with $\sigma$.
The, according to the amount, largest occurring angular propulsion velocity is here the same ($\omega=\SI{-0.12}{\second^{-1}}$) as in Fig.\ \ref{fig:fig3}c and thus, as explained in Section \ref{sec:propulsionL}, negligible compared to rotational Brownian motion.

\section{\label{conclusions}Conclusions}
We have simulated a cylindrical particle with concave and convex hemi-spherical ends, as has been used in many experiments \cite{WangCHM2012,AhmedEtAl2013,GarciaGradillaEtAl2013,GarciaGradillaSSKYWGW2014,WangLMAHM2014,AhmedGFM2014,BalkEtAl2014,WuEtAl2015a,EstebanFernandezdeAvilaMSLRCVMGZW2015,XuEtAl2015,Kiristi2015,EstebanEtAl2016,ZhouYWDW2017,ZhouZWW2017,EstebanFernandezEtAl2017,UygunEtAl2017,Zhou2018,HansenEtAl2018,WangGWSGXH2018,EstebanEtAl2018,BeltranEtAl2019,QualliotineEtAl2019,DumyJMBGMHA2020,McneillSWOLNM2021,AhmedWBGHM2016}, in the presence of a planar traveling ultrasound wave.
The particle has a variable cylinder length and diameter and includes (for a small cylinder length) hemi-spherical cups (nanoshells) \cite{SotoWGGGLKACW2016,TangEtAl2019,VossW2020} as a limiting case.   
On this basis, we have studied how the acoustic propulsion of such a bullet-shaped or cup-shaped particle depends on its length and diameter. 

Our results show a strong dependence of the propulsion on the dimensions of the particle. 
For example, we found that the flow field around the particle is dominated by an assembly of vortices whose positions move in a complex way when the particle length or the particle diameter is varied. 
Furthermore, we found that translational propulsion of the particle parallel to its symmetry axis changes sign at small diameters and, according to the amount, increases with both the length and diameter of the particle. 
Especially a large diameter allows to reach large propulsion speeds. 
The angular propulsion of the particle, on the other hand, decreases when the length or diameter is increased. 
This suggests that larger particles exhibit a stronger and more translational motion than smaller particles. 

These results are important since they contribute to a better understanding of the acoustic propulsion of nano- and microparticles, which will be a prerequisite for realizing the intriguing future applications that have been envisaged for such particles \cite{GarciaGradillaEtAl2013,GarciaGradillaSSKYWGW2014,WuEtAl2015a,EstebanFernandezdeAvilaMSLRCVMGZW2015,Kiristi2015,EstebanEtAl2016,ZhouYWDW2017,UygunEtAl2017,EstebanFernandezEtAl2017,HansenEtAl2018,EstebanEtAl2018,BeltranEtAl2019,QualliotineEtAl2019,Venugopalan2020}. The obtained insights into how the acoustic propulsion depends on the shape parameters of the particles will prospectively prove useful for selecting suitable particle designs for coming experimental studies. 
Although we have studied here two very common types of ultrasound-propelled particles, further studies that focus on other particle shapes will be required to complement the understanding of how acoustic propulsion depends on the particle shape.

\section*{Data availability}
The data that support the findings of this study are openly available in \url{https://doi.org/10.5281/zenodo.5913332}.

\section*{Conflicts of interest}
There are no conflicts of interest to declare.

\begin{acknowledgments}
We thank Patrick Kurzeja for helpful discussions. 
R.W.\ is funded by the Deutsche Forschungsgemeinschaft (DFG, German Research Foundation) -- WI 4170/3-1. 
The simulations for this work were performed on the computer cluster PALMA II of the University of M\"unster. 
\end{acknowledgments}

\nocite{apsrev41Control}
\bibliographystyle{apsrev4-1}
\bibliography{control,refs}

\begin{thebibliography}{76}%
\makeatletter
\providecommand \@ifxundefined [1]{%
 \@ifx{#1\undefined}
}%
\providecommand \@ifnum [1]{%
 \ifnum #1\expandafter \@firstoftwo
 \else \expandafter \@secondoftwo
 \fi
}%
\providecommand \@ifx [1]{%
 \ifx #1\expandafter \@firstoftwo
 \else \expandafter \@secondoftwo
 \fi
}%
\providecommand \natexlab [1]{#1}%
\providecommand \enquote  [1]{``#1''}%
\providecommand \bibnamefont  [1]{#1}%
\providecommand \bibfnamefont [1]{#1}%
\providecommand \citenamefont [1]{#1}%
\providecommand \href@noop [0]{\@secondoftwo}%
\providecommand \href [0]{\begingroup \@sanitize@url \@href}%
\providecommand \@href[1]{\@@startlink{#1}\@@href}%
\providecommand \@@href[1]{\endgroup#1\@@endlink}%
\providecommand \@sanitize@url [0]{\catcode `\\12\catcode `\$12\catcode
  `\&12\catcode `\#12\catcode `\^12\catcode `\_12\catcode `\%12\relax}%
\providecommand \@@startlink[1]{}%
\providecommand \@@endlink[0]{}%
\providecommand \url  [0]{\begingroup\@sanitize@url \@url }%
\providecommand \@url [1]{\endgroup\@href {#1}{\urlprefix }}%
\providecommand \urlprefix  [0]{URL }%
\providecommand \Eprint [0]{\href }%
\providecommand \doibase [0]{http://dx.doi.org/}%
\providecommand \selectlanguage [0]{\@gobble}%
\providecommand \bibinfo  [0]{\@secondoftwo}%
\providecommand \bibfield  [0]{\@secondoftwo}%
\providecommand \translation [1]{[#1]}%
\providecommand \BibitemOpen [0]{}%
\providecommand \bibitemStop [0]{}%
\providecommand \bibitemNoStop [0]{.\EOS\space}%
\providecommand \EOS [0]{\spacefactor3000\relax}%
\providecommand \BibitemShut  [1]{\csname bibitem#1\endcsname}%
\let\auto@bib@innerbib\@empty
\bibitem [{\citenamefont {Wang}\ \emph {et~al.}(2012)\citenamefont {Wang},
  \citenamefont {Castro}, \citenamefont {Hoyos},\ and\ \citenamefont
  {Mallouk}}]{WangCHM2012}%
  \BibitemOpen
  \bibfield  {author} {\bibinfo {author} {\bibfnamefont {W.}~\bibnamefont
  {Wang}}, \bibinfo {author} {\bibfnamefont {L.}~\bibnamefont {Castro}},
  \bibinfo {author} {\bibfnamefont {M.}~\bibnamefont {Hoyos}}, \ and\ \bibinfo
  {author} {\bibfnamefont {T.~E.}\ \bibnamefont {Mallouk}},\ }\bibfield
  {title} {\enquote {\bibinfo {title} {Autonomous motion of metallic microrods
  propelled by ultrasound},}\ }\href@noop {} {\bibfield  {journal} {\bibinfo
  {journal} {ACS Nano}\ }\textbf {\bibinfo {volume} {6}},\ \bibinfo {pages}
  {6122--6132} (\bibinfo {year} {2012})}\BibitemShut {NoStop}%
\bibitem [{\citenamefont {{Garcia-Gradilla}}\ \emph {et~al.}(2013)\citenamefont
  {{Garcia-Gradilla}}, \citenamefont {Orozco}, \citenamefont
  {Sattayasamitsathit}, \citenamefont {Soto}, \citenamefont {Kuralay},
  \citenamefont {Pourazary}, \citenamefont {Katzenberg}, \citenamefont {Gao},
  \citenamefont {Shen},\ and\ \citenamefont {Wang}}]{GarciaGradillaEtAl2013}%
  \BibitemOpen
  \bibfield  {author} {\bibinfo {author} {\bibfnamefont {V.}~\bibnamefont
  {{Garcia-Gradilla}}}, \bibinfo {author} {\bibfnamefont {J.}~\bibnamefont
  {Orozco}}, \bibinfo {author} {\bibfnamefont {S.}~\bibnamefont
  {Sattayasamitsathit}}, \bibinfo {author} {\bibfnamefont {F.}~\bibnamefont
  {Soto}}, \bibinfo {author} {\bibfnamefont {F.}~\bibnamefont {Kuralay}},
  \bibinfo {author} {\bibfnamefont {A.}~\bibnamefont {Pourazary}}, \bibinfo
  {author} {\bibfnamefont {A.}~\bibnamefont {Katzenberg}}, \bibinfo {author}
  {\bibfnamefont {W.}~\bibnamefont {Gao}}, \bibinfo {author} {\bibfnamefont
  {Y.}~\bibnamefont {Shen}}, \ and\ \bibinfo {author} {\bibfnamefont
  {J.}~\bibnamefont {Wang}},\ }\bibfield  {title} {\enquote {\bibinfo {title}
  {Functionalized ultrasound-propelled magnetically guided nanomotors: toward
  practical biomedical applications},}\ }\href@noop {} {\bibfield  {journal}
  {\bibinfo  {journal} {ACS Nano}\ }\textbf {\bibinfo {volume} {7}},\ \bibinfo
  {pages} {9232--9240} (\bibinfo {year} {2013})}\BibitemShut {NoStop}%
\bibitem [{\citenamefont {Ahmed}\ \emph {et~al.}(2013)\citenamefont {Ahmed},
  \citenamefont {Wang}, \citenamefont {Mair}, \citenamefont {Fraleigh},
  \citenamefont {Li}, \citenamefont {Castro}, \citenamefont {Hoyos},
  \citenamefont {Huang},\ and\ \citenamefont {Mallouk}}]{AhmedEtAl2013}%
  \BibitemOpen
  \bibfield  {author} {\bibinfo {author} {\bibfnamefont {S.}~\bibnamefont
  {Ahmed}}, \bibinfo {author} {\bibfnamefont {W.}~\bibnamefont {Wang}},
  \bibinfo {author} {\bibfnamefont {L.~O.}\ \bibnamefont {Mair}}, \bibinfo
  {author} {\bibfnamefont {R.~D.}\ \bibnamefont {Fraleigh}}, \bibinfo {author}
  {\bibfnamefont {S.}~\bibnamefont {Li}}, \bibinfo {author} {\bibfnamefont
  {L.~A.}\ \bibnamefont {Castro}}, \bibinfo {author} {\bibfnamefont
  {M.}~\bibnamefont {Hoyos}}, \bibinfo {author} {\bibfnamefont {T.~J.}\
  \bibnamefont {Huang}}, \ and\ \bibinfo {author} {\bibfnamefont {T.~E.}\
  \bibnamefont {Mallouk}},\ }\bibfield  {title} {\enquote {\bibinfo {title}
  {Steering acoustically propelled nanowire motors toward cells in a
  biologically compatible environment using magnetic fields},}\ }\href@noop {}
  {\bibfield  {journal} {\bibinfo  {journal} {Langmuir}\ }\textbf {\bibinfo
  {volume} {29}},\ \bibinfo {pages} {16113--16118} (\bibinfo {year}
  {2013})}\BibitemShut {NoStop}%
\bibitem [{\citenamefont {{Nadal}}\ and\ \citenamefont
  {{Lauga}}(2014)}]{NadalL2014}%
  \BibitemOpen
  \bibfield  {author} {\bibinfo {author} {\bibfnamefont {F.}~\bibnamefont
  {{Nadal}}}\ and\ \bibinfo {author} {\bibfnamefont {E.}~\bibnamefont
  {{Lauga}}},\ }\bibfield  {title} {\enquote {\bibinfo {title} {Asymmetric
  steady streaming as a mechanism for acoustic propulsion of rigid bodies},}\
  }\href@noop {} {\bibfield  {journal} {\bibinfo  {journal} {Physics of
  Fluids}\ }\textbf {\bibinfo {volume} {26}},\ \bibinfo {pages} {082001}
  (\bibinfo {year} {2014})}\BibitemShut {NoStop}%
\bibitem [{\citenamefont {Wu}\ \emph {et~al.}(2014)\citenamefont {Wu},
  \citenamefont {Li}, \citenamefont {Li}, \citenamefont {Gao}, \citenamefont
  {Xu}, \citenamefont {Christianson}, \citenamefont {Gao}, \citenamefont
  {Galarnyk}, \citenamefont {He}, \citenamefont {Zhang} \emph
  {et~al.}}]{WuEtAl2014}%
  \BibitemOpen
  \bibfield  {author} {\bibinfo {author} {\bibfnamefont {Z.}~\bibnamefont
  {Wu}}, \bibinfo {author} {\bibfnamefont {T.}~\bibnamefont {Li}}, \bibinfo
  {author} {\bibfnamefont {J.}~\bibnamefont {Li}}, \bibinfo {author}
  {\bibfnamefont {W.}~\bibnamefont {Gao}}, \bibinfo {author} {\bibfnamefont
  {T.}~\bibnamefont {Xu}}, \bibinfo {author} {\bibfnamefont {C.}~\bibnamefont
  {Christianson}}, \bibinfo {author} {\bibfnamefont {W.}~\bibnamefont {Gao}},
  \bibinfo {author} {\bibfnamefont {M.}~\bibnamefont {Galarnyk}}, \bibinfo
  {author} {\bibfnamefont {Q.}~\bibnamefont {He}}, \bibinfo {author}
  {\bibfnamefont {L.}~\bibnamefont {Zhang}},  \emph {et~al.},\ }\bibfield
  {title} {\enquote {\bibinfo {title} {Turning erythrocytes into functional
  micromotors},}\ }\href@noop {} {\bibfield  {journal} {\bibinfo  {journal}
  {ACS Nano}\ }\textbf {\bibinfo {volume} {8}},\ \bibinfo {pages}
  {12041--12048} (\bibinfo {year} {2014})}\BibitemShut {NoStop}%
\bibitem [{\citenamefont {Wang}\ \emph {et~al.}(2014)\citenamefont {Wang},
  \citenamefont {Li}, \citenamefont {Mair}, \citenamefont {Ahmed},
  \citenamefont {Huang},\ and\ \citenamefont {Mallouk}}]{WangLMAHM2014}%
  \BibitemOpen
  \bibfield  {author} {\bibinfo {author} {\bibfnamefont {W.}~\bibnamefont
  {Wang}}, \bibinfo {author} {\bibfnamefont {S.}~\bibnamefont {Li}}, \bibinfo
  {author} {\bibfnamefont {L.}~\bibnamefont {Mair}}, \bibinfo {author}
  {\bibfnamefont {S.}~\bibnamefont {Ahmed}}, \bibinfo {author} {\bibfnamefont
  {T.~J.}\ \bibnamefont {Huang}}, \ and\ \bibinfo {author} {\bibfnamefont
  {T.~E.}\ \bibnamefont {Mallouk}},\ }\bibfield  {title} {\enquote {\bibinfo
  {title} {Acoustic propulsion of nanorod motors inside living cells},}\
  }\href@noop {} {\bibfield  {journal} {\bibinfo  {journal} {Angewandte Chemie
  International Edition}\ }\textbf {\bibinfo {volume} {53}},\ \bibinfo {pages}
  {3201--3204} (\bibinfo {year} {2014})}\BibitemShut {NoStop}%
\bibitem [{\citenamefont {{Garcia-Gradilla}}\ \emph {et~al.}(2014)\citenamefont
  {{Garcia-Gradilla}}, \citenamefont {Sattayasamitsathit}, \citenamefont
  {Soto}, \citenamefont {Kuralay}, \citenamefont {Yard{\i}mc{\i}},
  \citenamefont {Wiitala}, \citenamefont {Galarnyk},\ and\ \citenamefont
  {Wang}}]{GarciaGradillaSSKYWGW2014}%
  \BibitemOpen
  \bibfield  {author} {\bibinfo {author} {\bibfnamefont {V.}~\bibnamefont
  {{Garcia-Gradilla}}}, \bibinfo {author} {\bibfnamefont {S.}~\bibnamefont
  {Sattayasamitsathit}}, \bibinfo {author} {\bibfnamefont {F.}~\bibnamefont
  {Soto}}, \bibinfo {author} {\bibfnamefont {F.}~\bibnamefont {Kuralay}},
  \bibinfo {author} {\bibfnamefont {C.}~\bibnamefont {Yard{\i}mc{\i}}},
  \bibinfo {author} {\bibfnamefont {D.}~\bibnamefont {Wiitala}}, \bibinfo
  {author} {\bibfnamefont {M.}~\bibnamefont {Galarnyk}}, \ and\ \bibinfo
  {author} {\bibfnamefont {J.}~\bibnamefont {Wang}},\ }\bibfield  {title}
  {\enquote {\bibinfo {title} {Ultrasound-propelled nanoporous gold wire for
  efficient drug loading and release},}\ }\href@noop {} {\bibfield  {journal}
  {\bibinfo  {journal} {Small}\ }\textbf {\bibinfo {volume} {10}},\ \bibinfo
  {pages} {4154--4159} (\bibinfo {year} {2014})}\BibitemShut {NoStop}%
\bibitem [{\citenamefont {Balk}\ \emph {et~al.}(2014)\citenamefont {Balk},
  \citenamefont {Mair}, \citenamefont {Mathai}, \citenamefont {Patrone},
  \citenamefont {Wang}, \citenamefont {Ahmed}, \citenamefont {Mallouk},
  \citenamefont {Liddle},\ and\ \citenamefont {Stavis}}]{BalkEtAl2014}%
  \BibitemOpen
  \bibfield  {author} {\bibinfo {author} {\bibfnamefont {A.~L.}\ \bibnamefont
  {Balk}}, \bibinfo {author} {\bibfnamefont {L.~O.}\ \bibnamefont {Mair}},
  \bibinfo {author} {\bibfnamefont {P.~P.}\ \bibnamefont {Mathai}}, \bibinfo
  {author} {\bibfnamefont {P.~N.}\ \bibnamefont {Patrone}}, \bibinfo {author}
  {\bibfnamefont {W.}~\bibnamefont {Wang}}, \bibinfo {author} {\bibfnamefont
  {S.}~\bibnamefont {Ahmed}}, \bibinfo {author} {\bibfnamefont {T.~E.}\
  \bibnamefont {Mallouk}}, \bibinfo {author} {\bibfnamefont {J.~A.}\
  \bibnamefont {Liddle}}, \ and\ \bibinfo {author} {\bibfnamefont {S.~M.}\
  \bibnamefont {Stavis}},\ }\bibfield  {title} {\enquote {\bibinfo {title}
  {Kilohertz rotation of nanorods propelled by ultrasound, traced by
  microvortex advection of nanoparticles},}\ }\href@noop {} {\bibfield
  {journal} {\bibinfo  {journal} {ACS Nano}\ }\textbf {\bibinfo {volume} {8}},\
  \bibinfo {pages} {8300--8309} (\bibinfo {year} {2014})}\BibitemShut {NoStop}%
\bibitem [{\citenamefont {Ahmed}\ \emph {et~al.}(2014)\citenamefont {Ahmed},
  \citenamefont {Gentekos}, \citenamefont {Fink},\ and\ \citenamefont
  {Mallouk}}]{AhmedGFM2014}%
  \BibitemOpen
  \bibfield  {author} {\bibinfo {author} {\bibfnamefont {S.}~\bibnamefont
  {Ahmed}}, \bibinfo {author} {\bibfnamefont {D.~T.}\ \bibnamefont {Gentekos}},
  \bibinfo {author} {\bibfnamefont {C.~A.}\ \bibnamefont {Fink}}, \ and\
  \bibinfo {author} {\bibfnamefont {T.~E.}\ \bibnamefont {Mallouk}},\
  }\bibfield  {title} {\enquote {\bibinfo {title} {Self-assembly of nanorod
  motors into geometrically regular multimers and their propulsion by
  ultrasound},}\ }\href@noop {} {\bibfield  {journal} {\bibinfo  {journal} {ACS
  Nano}\ }\textbf {\bibinfo {volume} {8}},\ \bibinfo {pages} {11053--11060}
  (\bibinfo {year} {2014})}\BibitemShut {NoStop}%
\bibitem [{\citenamefont {Ahmed}\ \emph {et~al.}(2015)\citenamefont {Ahmed},
  \citenamefont {Lu}, \citenamefont {Nourhani}, \citenamefont {Lammert},
  \citenamefont {Stratton}, \citenamefont {Muddana}, \citenamefont {Crespi},\
  and\ \citenamefont {Huang}}]{AhmedLNLSMCH2015}%
  \BibitemOpen
  \bibfield  {author} {\bibinfo {author} {\bibfnamefont {D.}~\bibnamefont
  {Ahmed}}, \bibinfo {author} {\bibfnamefont {M.}~\bibnamefont {Lu}}, \bibinfo
  {author} {\bibfnamefont {A.}~\bibnamefont {Nourhani}}, \bibinfo {author}
  {\bibfnamefont {P.~E.}\ \bibnamefont {Lammert}}, \bibinfo {author}
  {\bibfnamefont {Z.}~\bibnamefont {Stratton}}, \bibinfo {author}
  {\bibfnamefont {H.~S.}\ \bibnamefont {Muddana}}, \bibinfo {author}
  {\bibfnamefont {V.~H.}\ \bibnamefont {Crespi}}, \ and\ \bibinfo {author}
  {\bibfnamefont {T.~J.}\ \bibnamefont {Huang}},\ }\bibfield  {title} {\enquote
  {\bibinfo {title} {Selectively manipulable acoustic-powered microswimmers},}\
  }\href@noop {} {\bibfield  {journal} {\bibinfo  {journal} {Scientific
  Reports}\ }\textbf {\bibinfo {volume} {5}},\ \bibinfo {pages} {9744}
  (\bibinfo {year} {2015})}\BibitemShut {NoStop}%
\bibitem [{\citenamefont {Wang}\ \emph {et~al.}(2015)\citenamefont {Wang},
  \citenamefont {Duan}, \citenamefont {Zhang}, \citenamefont {Sun},
  \citenamefont {Sen},\ and\ \citenamefont {Mallouk}}]{WangDZSSM2015}%
  \BibitemOpen
  \bibfield  {author} {\bibinfo {author} {\bibfnamefont {W.}~\bibnamefont
  {Wang}}, \bibinfo {author} {\bibfnamefont {W.}~\bibnamefont {Duan}}, \bibinfo
  {author} {\bibfnamefont {Z.}~\bibnamefont {Zhang}}, \bibinfo {author}
  {\bibfnamefont {M.}~\bibnamefont {Sun}}, \bibinfo {author} {\bibfnamefont
  {A.}~\bibnamefont {Sen}}, \ and\ \bibinfo {author} {\bibfnamefont {T.~E.}\
  \bibnamefont {Mallouk}},\ }\bibfield  {title} {\enquote {\bibinfo {title} {A
  tale of two forces: simultaneous chemical and acoustic propulsion of
  bimetallic micromotors},}\ }\href@noop {} {\bibfield  {journal} {\bibinfo
  {journal} {Chemical Communications}\ }\textbf {\bibinfo {volume} {51}},\
  \bibinfo {pages} {1020--1023} (\bibinfo {year} {2015})}\BibitemShut {NoStop}%
\bibitem [{\citenamefont {{Esteban-Fern{\'a}ndez de {\'A}vila}}\ \emph
  {et~al.}(2015)\citenamefont {{Esteban-Fern{\'a}ndez de {\'A}vila}},
  \citenamefont {Mart{\'\i}n}, \citenamefont {Soto}, \citenamefont
  {{Lopez-Ramirez}}, \citenamefont {Campuzano}, \citenamefont
  {{V{\'a}squez-Machado}}, \citenamefont {Gao}, \citenamefont {Zhang},\ and\
  \citenamefont {Wang}}]{EstebanFernandezdeAvilaMSLRCVMGZW2015}%
  \BibitemOpen
  \bibfield  {author} {\bibinfo {author} {\bibfnamefont {B.}~\bibnamefont
  {{Esteban-Fern{\'a}ndez de {\'A}vila}}}, \bibinfo {author} {\bibfnamefont
  {A.}~\bibnamefont {Mart{\'\i}n}}, \bibinfo {author} {\bibfnamefont
  {F.}~\bibnamefont {Soto}}, \bibinfo {author} {\bibfnamefont {M.~A.}\
  \bibnamefont {{Lopez-Ramirez}}}, \bibinfo {author} {\bibfnamefont
  {S.}~\bibnamefont {Campuzano}}, \bibinfo {author} {\bibfnamefont {G.~M.}\
  \bibnamefont {{V{\'a}squez-Machado}}}, \bibinfo {author} {\bibfnamefont
  {W.}~\bibnamefont {Gao}}, \bibinfo {author} {\bibfnamefont {L.}~\bibnamefont
  {Zhang}}, \ and\ \bibinfo {author} {\bibfnamefont {J.}~\bibnamefont {Wang}},\
  }\bibfield  {title} {\enquote {\bibinfo {title} {Single cell real-time
  mi{RNA}s sensing based on nanomotors},}\ }\href@noop {} {\bibfield  {journal}
  {\bibinfo  {journal} {ACS Nano}\ }\textbf {\bibinfo {volume} {9}},\ \bibinfo
  {pages} {6756--6764} (\bibinfo {year} {2015})}\BibitemShut {NoStop}%
\bibitem [{\citenamefont {Wu}\ \emph {et~al.}(2015{\natexlab{a}})\citenamefont
  {Wu}, \citenamefont {Li}, \citenamefont {Gao}, \citenamefont {Xu},
  \citenamefont {{Jurado-S{\'a}nchez}}, \citenamefont {Li}, \citenamefont
  {Gao}, \citenamefont {He}, \citenamefont {Zhang},\ and\ \citenamefont
  {Wang}}]{WuEtAl2015a}%
  \BibitemOpen
  \bibfield  {author} {\bibinfo {author} {\bibfnamefont {Z.}~\bibnamefont
  {Wu}}, \bibinfo {author} {\bibfnamefont {T.}~\bibnamefont {Li}}, \bibinfo
  {author} {\bibfnamefont {W.}~\bibnamefont {Gao}}, \bibinfo {author}
  {\bibfnamefont {W.}~\bibnamefont {Xu}}, \bibinfo {author} {\bibfnamefont
  {B.}~\bibnamefont {{Jurado-S{\'a}nchez}}}, \bibinfo {author} {\bibfnamefont
  {J.}~\bibnamefont {Li}}, \bibinfo {author} {\bibfnamefont {W.}~\bibnamefont
  {Gao}}, \bibinfo {author} {\bibfnamefont {Q.}~\bibnamefont {He}}, \bibinfo
  {author} {\bibfnamefont {L.}~\bibnamefont {Zhang}}, \ and\ \bibinfo {author}
  {\bibfnamefont {J.}~\bibnamefont {Wang}},\ }\bibfield  {title} {\enquote
  {\bibinfo {title} {Cell-membrane-coated synthetic nanomotors for effective
  biodetoxification},}\ }\href@noop {} {\bibfield  {journal} {\bibinfo
  {journal} {Advanced Functional Materials}\ }\textbf {\bibinfo {volume}
  {25}},\ \bibinfo {pages} {3881--3887} (\bibinfo {year}
  {2015}{\natexlab{a}})}\BibitemShut {NoStop}%
\bibitem [{\citenamefont {Wu}\ \emph {et~al.}(2015{\natexlab{b}})\citenamefont
  {Wu}, \citenamefont {{Esteban-Fern{\'a}ndez de {\'A}vila}}, \citenamefont
  {Mart{\'i}n}, \citenamefont {Christianson}, \citenamefont {Gao},
  \citenamefont {Thamphiwatana}, \citenamefont {Escarpa}, \citenamefont {He},
  \citenamefont {Zhang},\ and\ \citenamefont {Wang}}]{WuEtAl2015b}%
  \BibitemOpen
  \bibfield  {author} {\bibinfo {author} {\bibfnamefont {Z.}~\bibnamefont
  {Wu}}, \bibinfo {author} {\bibfnamefont {B.}~\bibnamefont
  {{Esteban-Fern{\'a}ndez de {\'A}vila}}}, \bibinfo {author} {\bibfnamefont
  {A.}~\bibnamefont {Mart{\'i}n}}, \bibinfo {author} {\bibfnamefont
  {C.}~\bibnamefont {Christianson}}, \bibinfo {author} {\bibfnamefont
  {W.}~\bibnamefont {Gao}}, \bibinfo {author} {\bibfnamefont {S.~K.}\
  \bibnamefont {Thamphiwatana}}, \bibinfo {author} {\bibfnamefont
  {A.}~\bibnamefont {Escarpa}}, \bibinfo {author} {\bibfnamefont
  {Q.}~\bibnamefont {He}}, \bibinfo {author} {\bibfnamefont {L.}~\bibnamefont
  {Zhang}}, \ and\ \bibinfo {author} {\bibfnamefont {J.}~\bibnamefont {Wang}},\
  }\bibfield  {title} {\enquote {\bibinfo {title} {{RBC} micromotors carrying
  multiple cargos towards potential theranostic applications},}\ }\href@noop {}
  {\bibfield  {journal} {\bibinfo  {journal} {Nanoscale}\ }\textbf {\bibinfo
  {volume} {7}},\ \bibinfo {pages} {13680--13686} (\bibinfo {year}
  {2015}{\natexlab{b}})}\BibitemShut {NoStop}%
\bibitem [{\citenamefont {Rao}\ \emph {et~al.}(2015)\citenamefont {Rao},
  \citenamefont {Li}, \citenamefont {Meng}, \citenamefont {Zheng},
  \citenamefont {Cai},\ and\ \citenamefont {Wang}}]{RaoLMZCW2015}%
  \BibitemOpen
  \bibfield  {author} {\bibinfo {author} {\bibfnamefont {K.~J.}\ \bibnamefont
  {Rao}}, \bibinfo {author} {\bibfnamefont {F.}~\bibnamefont {Li}}, \bibinfo
  {author} {\bibfnamefont {L.}~\bibnamefont {Meng}}, \bibinfo {author}
  {\bibfnamefont {H.}~\bibnamefont {Zheng}}, \bibinfo {author} {\bibfnamefont
  {F.}~\bibnamefont {Cai}}, \ and\ \bibinfo {author} {\bibfnamefont
  {W.}~\bibnamefont {Wang}},\ }\bibfield  {title} {\enquote {\bibinfo {title}
  {A force to be reckoned with: a review of synthetic microswimmers powered by
  ultrasound},}\ }\href@noop {} {\bibfield  {journal} {\bibinfo  {journal}
  {Small}\ }\textbf {\bibinfo {volume} {11}},\ \bibinfo {pages} {2836--2846}
  (\bibinfo {year} {2015})}\BibitemShut {NoStop}%
\bibitem [{\citenamefont {Kiristi}\ \emph {et~al.}(2015)\citenamefont
  {Kiristi}, \citenamefont {Singh}, \citenamefont {{Esteban-Fern{\'a}ndez de
  {\'A}vila}}, \citenamefont {Uygun}, \citenamefont {Soto}, \citenamefont
  {{Aktas Uygun}},\ and\ \citenamefont {Wang}}]{Kiristi2015}%
  \BibitemOpen
  \bibfield  {author} {\bibinfo {author} {\bibfnamefont {M.}~\bibnamefont
  {Kiristi}}, \bibinfo {author} {\bibfnamefont {V.}~\bibnamefont {Singh}},
  \bibinfo {author} {\bibfnamefont {B.}~\bibnamefont {{Esteban-Fern{\'a}ndez de
  {\'A}vila}}}, \bibinfo {author} {\bibfnamefont {M.}~\bibnamefont {Uygun}},
  \bibinfo {author} {\bibfnamefont {F.}~\bibnamefont {Soto}}, \bibinfo {author}
  {\bibfnamefont {D.}~\bibnamefont {{Aktas Uygun}}}, \ and\ \bibinfo {author}
  {\bibfnamefont {J.}~\bibnamefont {Wang}},\ }\bibfield  {title} {\enquote
  {\bibinfo {title} {Lysozyme-based antibacterial nanomotors},}\ }\href@noop {}
  {\bibfield  {journal} {\bibinfo  {journal} {ACS Nano}\ }\textbf {\bibinfo
  {volume} {9}},\ \bibinfo {pages} {9252--9259} (\bibinfo {year}
  {2015})}\BibitemShut {NoStop}%
\bibitem [{\citenamefont {Kim}\ \emph {et~al.}(2016)\citenamefont {Kim},
  \citenamefont {Guo}, \citenamefont {Liang}, \citenamefont {Zhu},\ and\
  \citenamefont {Fan}}]{KimGLZF2016}%
  \BibitemOpen
  \bibfield  {author} {\bibinfo {author} {\bibfnamefont {K.}~\bibnamefont
  {Kim}}, \bibinfo {author} {\bibfnamefont {J.}~\bibnamefont {Guo}}, \bibinfo
  {author} {\bibfnamefont {Z.}~\bibnamefont {Liang}}, \bibinfo {author}
  {\bibfnamefont {F.}~\bibnamefont {Zhu}}, \ and\ \bibinfo {author}
  {\bibfnamefont {D.}~\bibnamefont {Fan}},\ }\bibfield  {title} {\enquote
  {\bibinfo {title} {Man-made rotary nanomotors: a review of recent
  developments},}\ }\href@noop {} {\bibfield  {journal} {\bibinfo  {journal}
  {Nanoscale}\ }\textbf {\bibinfo {volume} {8}},\ \bibinfo {pages}
  {10471--10490} (\bibinfo {year} {2016})}\BibitemShut {NoStop}%
\bibitem [{\citenamefont {{Esteban-Fern{\'a}ndez de {\'A}vila}}\ \emph
  {et~al.}(2016)\citenamefont {{Esteban-Fern{\'a}ndez de {\'A}vila}},
  \citenamefont {Angell}, \citenamefont {Soto}, \citenamefont
  {{Lopez-Ramirez}}, \citenamefont {B{\'a}ez}, \citenamefont {Xie},
  \citenamefont {Wang},\ and\ \citenamefont {Chen}}]{EstebanEtAl2016}%
  \BibitemOpen
  \bibfield  {author} {\bibinfo {author} {\bibfnamefont {B.}~\bibnamefont
  {{Esteban-Fern{\'a}ndez de {\'A}vila}}}, \bibinfo {author} {\bibfnamefont
  {C.}~\bibnamefont {Angell}}, \bibinfo {author} {\bibfnamefont
  {F.}~\bibnamefont {Soto}}, \bibinfo {author} {\bibfnamefont {M.~A.}\
  \bibnamefont {{Lopez-Ramirez}}}, \bibinfo {author} {\bibfnamefont {D.~F.}\
  \bibnamefont {B{\'a}ez}}, \bibinfo {author} {\bibfnamefont {S.}~\bibnamefont
  {Xie}}, \bibinfo {author} {\bibfnamefont {J.}~\bibnamefont {Wang}}, \ and\
  \bibinfo {author} {\bibfnamefont {Y.}~\bibnamefont {Chen}},\ }\bibfield
  {title} {\enquote {\bibinfo {title} {Acoustically propelled nanomotors for
  intracellular si{RNA} delivery},}\ }\href@noop {} {\bibfield  {journal}
  {\bibinfo  {journal} {ACS Nano}\ }\textbf {\bibinfo {volume} {10}},\ \bibinfo
  {pages} {4997--5005} (\bibinfo {year} {2016})}\BibitemShut {NoStop}%
\bibitem [{\citenamefont {Soto}\ \emph {et~al.}(2016)\citenamefont {Soto},
  \citenamefont {Wagner}, \citenamefont {{Garcia-Gradilla}}, \citenamefont
  {Gillespie}, \citenamefont {Lakshmipathy}, \citenamefont {Karshalev},
  \citenamefont {Angell}, \citenamefont {Chen},\ and\ \citenamefont
  {Wang}}]{SotoWGGGLKACW2016}%
  \BibitemOpen
  \bibfield  {author} {\bibinfo {author} {\bibfnamefont {F.}~\bibnamefont
  {Soto}}, \bibinfo {author} {\bibfnamefont {G.~L.}\ \bibnamefont {Wagner}},
  \bibinfo {author} {\bibfnamefont {V.}~\bibnamefont {{Garcia-Gradilla}}},
  \bibinfo {author} {\bibfnamefont {K.~T.}\ \bibnamefont {Gillespie}}, \bibinfo
  {author} {\bibfnamefont {D.~R.}\ \bibnamefont {Lakshmipathy}}, \bibinfo
  {author} {\bibfnamefont {E.}~\bibnamefont {Karshalev}}, \bibinfo {author}
  {\bibfnamefont {C.}~\bibnamefont {Angell}}, \bibinfo {author} {\bibfnamefont
  {Y.}~\bibnamefont {Chen}}, \ and\ \bibinfo {author} {\bibfnamefont
  {J.}~\bibnamefont {Wang}},\ }\bibfield  {title} {\enquote {\bibinfo {title}
  {Acoustically propelled nanoshells},}\ }\href@noop {} {\bibfield  {journal}
  {\bibinfo  {journal} {Nanoscale}\ }\textbf {\bibinfo {volume} {8}},\ \bibinfo
  {pages} {17788--17793} (\bibinfo {year} {2016})}\BibitemShut {NoStop}%
\bibitem [{\citenamefont {Ahmed}\ \emph
  {et~al.}(2016{\natexlab{a}})\citenamefont {Ahmed}, \citenamefont {Wang},
  \citenamefont {Bai}, \citenamefont {Gentekos}, \citenamefont {Hoyos},\ and\
  \citenamefont {Mallouk}}]{AhmedWBGHM2016}%
  \BibitemOpen
  \bibfield  {author} {\bibinfo {author} {\bibfnamefont {S.}~\bibnamefont
  {Ahmed}}, \bibinfo {author} {\bibfnamefont {W.}~\bibnamefont {Wang}},
  \bibinfo {author} {\bibfnamefont {L.}~\bibnamefont {Bai}}, \bibinfo {author}
  {\bibfnamefont {D.~T.}\ \bibnamefont {Gentekos}}, \bibinfo {author}
  {\bibfnamefont {M.}~\bibnamefont {Hoyos}}, \ and\ \bibinfo {author}
  {\bibfnamefont {T.~E.}\ \bibnamefont {Mallouk}},\ }\bibfield  {title}
  {\enquote {\bibinfo {title} {Density and shape effects in the acoustic
  propulsion of bimetallic nanorod motors},}\ }\href@noop {} {\bibfield
  {journal} {\bibinfo  {journal} {ACS Nano}\ }\textbf {\bibinfo {volume}
  {10}},\ \bibinfo {pages} {4763--4769} (\bibinfo {year}
  {2016}{\natexlab{a}})}\BibitemShut {NoStop}%
\bibitem [{\citenamefont {Kaynak}\ \emph {et~al.}(2016)\citenamefont {Kaynak},
  \citenamefont {Ozcelik}, \citenamefont {Nama}, \citenamefont {Nourhani},
  \citenamefont {Lammert}, \citenamefont {Crespi},\ and\ \citenamefont
  {Huang}}]{KaynakONNLCH2016}%
  \BibitemOpen
  \bibfield  {author} {\bibinfo {author} {\bibfnamefont {M.}~\bibnamefont
  {Kaynak}}, \bibinfo {author} {\bibfnamefont {A.}~\bibnamefont {Ozcelik}},
  \bibinfo {author} {\bibfnamefont {N.}~\bibnamefont {Nama}}, \bibinfo {author}
  {\bibfnamefont {A.}~\bibnamefont {Nourhani}}, \bibinfo {author}
  {\bibfnamefont {P.~E.}\ \bibnamefont {Lammert}}, \bibinfo {author}
  {\bibfnamefont {V.~H.}\ \bibnamefont {Crespi}}, \ and\ \bibinfo {author}
  {\bibfnamefont {T.~J.}\ \bibnamefont {Huang}},\ }\bibfield  {title} {\enquote
  {\bibinfo {title} {Acoustofluidic actuation of in situ fabricated
  microrotors},}\ }\href@noop {} {\bibfield  {journal} {\bibinfo  {journal}
  {Lab on a Chip}\ }\textbf {\bibinfo {volume} {16}},\ \bibinfo {pages}
  {3532--3537} (\bibinfo {year} {2016})}\BibitemShut {NoStop}%
\bibitem [{\citenamefont {Uygun}\ \emph {et~al.}(2017)\citenamefont {Uygun},
  \citenamefont {{Jurado-S{\'a}nchez}}, \citenamefont {Uygun}, \citenamefont
  {Singh}, \citenamefont {Zhang},\ and\ \citenamefont {Wang}}]{UygunEtAl2017}%
  \BibitemOpen
  \bibfield  {author} {\bibinfo {author} {\bibfnamefont {M.}~\bibnamefont
  {Uygun}}, \bibinfo {author} {\bibfnamefont {B.}~\bibnamefont
  {{Jurado-S{\'a}nchez}}}, \bibinfo {author} {\bibfnamefont {D.~A.}\
  \bibnamefont {Uygun}}, \bibinfo {author} {\bibfnamefont {V.~V.}\ \bibnamefont
  {Singh}}, \bibinfo {author} {\bibfnamefont {L.}~\bibnamefont {Zhang}}, \ and\
  \bibinfo {author} {\bibfnamefont {J.}~\bibnamefont {Wang}},\ }\bibfield
  {title} {\enquote {\bibinfo {title} {Ultrasound-propelled nanowire motors
  enhance asparaginase enzymatic activity against cancer cells},}\ }\href@noop
  {} {\bibfield  {journal} {\bibinfo  {journal} {Nanoscale}\ }\textbf {\bibinfo
  {volume} {9}},\ \bibinfo {pages} {18423--18429} (\bibinfo {year}
  {2017})}\BibitemShut {NoStop}%
\bibitem [{\citenamefont {{Esteban-Fern{\'a}ndez de {\'A}vila}}\ \emph
  {et~al.}(2017)\citenamefont {{Esteban-Fern{\'a}ndez de {\'A}vila}},
  \citenamefont {{Ram{\'i}rez-Herrera}}, \citenamefont {Campuzano},
  \citenamefont {Angsantikul}, \citenamefont {Zhang},\ and\ \citenamefont
  {Wang}}]{EstebanFernandezEtAl2017}%
  \BibitemOpen
  \bibfield  {author} {\bibinfo {author} {\bibfnamefont {B.}~\bibnamefont
  {{Esteban-Fern{\'a}ndez de {\'A}vila}}}, \bibinfo {author} {\bibfnamefont
  {D.~E.}\ \bibnamefont {{Ram{\'i}rez-Herrera}}}, \bibinfo {author}
  {\bibfnamefont {S.}~\bibnamefont {Campuzano}}, \bibinfo {author}
  {\bibfnamefont {P.}~\bibnamefont {Angsantikul}}, \bibinfo {author}
  {\bibfnamefont {L.}~\bibnamefont {Zhang}}, \ and\ \bibinfo {author}
  {\bibfnamefont {J.}~\bibnamefont {Wang}},\ }\bibfield  {title} {\enquote
  {\bibinfo {title} {Nanomotor-enabled p{H}-responsive intracellular delivery
  of caspase-3: toward rapid cell apoptosis},}\ }\href@noop {} {\bibfield
  {journal} {\bibinfo  {journal} {ACS Nano}\ }\textbf {\bibinfo {volume}
  {11}},\ \bibinfo {pages} {5367--5374} (\bibinfo {year} {2017})}\BibitemShut
  {NoStop}%
\bibitem [{\citenamefont {Ren}\ \emph {et~al.}(2017)\citenamefont {Ren},
  \citenamefont {Zhou}, \citenamefont {Mao}, \citenamefont {Xu}, \citenamefont
  {Huang},\ and\ \citenamefont {Mallouk}}]{RenZMXHM2017}%
  \BibitemOpen
  \bibfield  {author} {\bibinfo {author} {\bibfnamefont {L.}~\bibnamefont
  {Ren}}, \bibinfo {author} {\bibfnamefont {D.}~\bibnamefont {Zhou}}, \bibinfo
  {author} {\bibfnamefont {Z.}~\bibnamefont {Mao}}, \bibinfo {author}
  {\bibfnamefont {P.}~\bibnamefont {Xu}}, \bibinfo {author} {\bibfnamefont
  {T.~J.}\ \bibnamefont {Huang}}, \ and\ \bibinfo {author} {\bibfnamefont
  {T.~E.}\ \bibnamefont {Mallouk}},\ }\bibfield  {title} {\enquote {\bibinfo
  {title} {Rheotaxis of bimetallic micromotors driven by chemical-acoustic
  hybrid power},}\ }\href@noop {} {\bibfield  {journal} {\bibinfo  {journal}
  {ACS Nano}\ }\textbf {\bibinfo {volume} {11}},\ \bibinfo {pages}
  {10591--10598} (\bibinfo {year} {2017})}\BibitemShut {NoStop}%
\bibitem [{\citenamefont {Kaynak}\ \emph {et~al.}(2017)\citenamefont {Kaynak},
  \citenamefont {Ozcelik}, \citenamefont {Nourhani}, \citenamefont {Lammert},
  \citenamefont {Crespi},\ and\ \citenamefont {Huang}}]{KaynakONLCH2017}%
  \BibitemOpen
  \bibfield  {author} {\bibinfo {author} {\bibfnamefont {M.}~\bibnamefont
  {Kaynak}}, \bibinfo {author} {\bibfnamefont {A.}~\bibnamefont {Ozcelik}},
  \bibinfo {author} {\bibfnamefont {A.}~\bibnamefont {Nourhani}}, \bibinfo
  {author} {\bibfnamefont {P.~E.}\ \bibnamefont {Lammert}}, \bibinfo {author}
  {\bibfnamefont {V.~H.}\ \bibnamefont {Crespi}}, \ and\ \bibinfo {author}
  {\bibfnamefont {T.~J.}\ \bibnamefont {Huang}},\ }\bibfield  {title} {\enquote
  {\bibinfo {title} {Acoustic actuation of bioinspired microswimmers},}\
  }\href@noop {} {\bibfield  {journal} {\bibinfo  {journal} {Lab on a Chip}\
  }\textbf {\bibinfo {volume} {17}},\ \bibinfo {pages} {395--400} (\bibinfo
  {year} {2017})}\BibitemShut {NoStop}%
\bibitem [{\citenamefont {Collis}\ \emph {et~al.}(2017)\citenamefont {Collis},
  \citenamefont {Chakraborty},\ and\ \citenamefont {Sader}}]{CollisCS2017}%
  \BibitemOpen
  \bibfield  {author} {\bibinfo {author} {\bibfnamefont {J.~F.}\ \bibnamefont
  {Collis}}, \bibinfo {author} {\bibfnamefont {D.}~\bibnamefont {Chakraborty}},
  \ and\ \bibinfo {author} {\bibfnamefont {J.~E.}\ \bibnamefont {Sader}},\
  }\bibfield  {title} {\enquote {\bibinfo {title} {Autonomous propulsion of
  nanorods trapped in an acoustic field},}\ }\href@noop {} {\bibfield
  {journal} {\bibinfo  {journal} {Journal of Fluid Mechanics}\ }\textbf
  {\bibinfo {volume} {825}},\ \bibinfo {pages} {29--48} (\bibinfo {year}
  {2017})}\BibitemShut {NoStop}%
\bibitem [{\citenamefont {Zhou}\ \emph
  {et~al.}(2017{\natexlab{a}})\citenamefont {Zhou}, \citenamefont {Zhao},
  \citenamefont {Wei},\ and\ \citenamefont {Wang}}]{ZhouZWW2017}%
  \BibitemOpen
  \bibfield  {author} {\bibinfo {author} {\bibfnamefont {C.}~\bibnamefont
  {Zhou}}, \bibinfo {author} {\bibfnamefont {L.}~\bibnamefont {Zhao}}, \bibinfo
  {author} {\bibfnamefont {M.}~\bibnamefont {Wei}}, \ and\ \bibinfo {author}
  {\bibfnamefont {W.}~\bibnamefont {Wang}},\ }\bibfield  {title} {\enquote
  {\bibinfo {title} {Twists and turns of orbiting and spinning metallic
  microparticles powered by megahertz ultrasound},}\ }\href@noop {} {\bibfield
  {journal} {\bibinfo  {journal} {ACS Nano}\ }\textbf {\bibinfo {volume}
  {11}},\ \bibinfo {pages} {12668--12676} (\bibinfo {year}
  {2017}{\natexlab{a}})}\BibitemShut {NoStop}%
\bibitem [{\citenamefont {Zhou}\ \emph
  {et~al.}(2017{\natexlab{b}})\citenamefont {Zhou}, \citenamefont {Yin},
  \citenamefont {Wu}, \citenamefont {Du},\ and\ \citenamefont
  {Wang}}]{ZhouYWDW2017}%
  \BibitemOpen
  \bibfield  {author} {\bibinfo {author} {\bibfnamefont {C.}~\bibnamefont
  {Zhou}}, \bibinfo {author} {\bibfnamefont {J.}~\bibnamefont {Yin}}, \bibinfo
  {author} {\bibfnamefont {C.}~\bibnamefont {Wu}}, \bibinfo {author}
  {\bibfnamefont {L.}~\bibnamefont {Du}}, \ and\ \bibinfo {author}
  {\bibfnamefont {Y.}~\bibnamefont {Wang}},\ }\bibfield  {title} {\enquote
  {\bibinfo {title} {Efficient target capture and transport by fuel-free
  micromotors in a multichannel microchip},}\ }\href@noop {} {\bibfield
  {journal} {\bibinfo  {journal} {Soft Matter}\ }\textbf {\bibinfo {volume}
  {13}},\ \bibinfo {pages} {8064--8069} (\bibinfo {year}
  {2017}{\natexlab{b}})}\BibitemShut {NoStop}%
\bibitem [{\citenamefont {Chen}\ \emph {et~al.}(2018)\citenamefont {Chen},
  \citenamefont {Jang}, \citenamefont {Ahmed}, \citenamefont {Hu},
  \citenamefont {{De Marco}}, \citenamefont {Hoop}, \citenamefont {Mushtaq},
  \citenamefont {Nelson},\ and\ \citenamefont {Pan{\'e}}}]{ChenEtAl2018}%
  \BibitemOpen
  \bibfield  {author} {\bibinfo {author} {\bibfnamefont {X.-Z.}\ \bibnamefont
  {Chen}}, \bibinfo {author} {\bibfnamefont {B.}~\bibnamefont {Jang}}, \bibinfo
  {author} {\bibfnamefont {D.}~\bibnamefont {Ahmed}}, \bibinfo {author}
  {\bibfnamefont {C.}~\bibnamefont {Hu}}, \bibinfo {author} {\bibfnamefont
  {C.}~\bibnamefont {{De Marco}}}, \bibinfo {author} {\bibfnamefont
  {M.}~\bibnamefont {Hoop}}, \bibinfo {author} {\bibfnamefont {F.}~\bibnamefont
  {Mushtaq}}, \bibinfo {author} {\bibfnamefont {B.~J.}\ \bibnamefont {Nelson}},
  \ and\ \bibinfo {author} {\bibfnamefont {S.}~\bibnamefont {Pan{\'e}}},\
  }\bibfield  {title} {\enquote {\bibinfo {title} {Small-scale machines driven
  by external power sources},}\ }\href@noop {} {\bibfield  {journal} {\bibinfo
  {journal} {Advanced Materials}\ }\textbf {\bibinfo {volume} {30}},\ \bibinfo
  {pages} {1705061} (\bibinfo {year} {2018})}\BibitemShut {NoStop}%
\bibitem [{\citenamefont {{Hansen-Bruhn}}\ \emph {et~al.}(2018)\citenamefont
  {{Hansen-Bruhn}}, \citenamefont {{Esteban-Fern{\'a}ndez de {\'A}vila}},
  \citenamefont {{Beltr{\'a}n-Gast{\'e}lum}}, \citenamefont {Zhao},
  \citenamefont {{Ram{\'\i}rez-Herrera}}, \citenamefont {Angsantikul},
  \citenamefont {{Vesterager Gothelf}}, \citenamefont {Zhang},\ and\
  \citenamefont {Wang}}]{HansenEtAl2018}%
  \BibitemOpen
  \bibfield  {author} {\bibinfo {author} {\bibfnamefont {M.}~\bibnamefont
  {{Hansen-Bruhn}}}, \bibinfo {author} {\bibfnamefont {B.}~\bibnamefont
  {{Esteban-Fern{\'a}ndez de {\'A}vila}}}, \bibinfo {author} {\bibfnamefont
  {M.}~\bibnamefont {{Beltr{\'a}n-Gast{\'e}lum}}}, \bibinfo {author}
  {\bibfnamefont {J.}~\bibnamefont {Zhao}}, \bibinfo {author} {\bibfnamefont
  {D.~E.}\ \bibnamefont {{Ram{\'\i}rez-Herrera}}}, \bibinfo {author}
  {\bibfnamefont {P.}~\bibnamefont {Angsantikul}}, \bibinfo {author}
  {\bibfnamefont {K.}~\bibnamefont {{Vesterager Gothelf}}}, \bibinfo {author}
  {\bibfnamefont {L.}~\bibnamefont {Zhang}}, \ and\ \bibinfo {author}
  {\bibfnamefont {J.}~\bibnamefont {Wang}},\ }\bibfield  {title} {\enquote
  {\bibinfo {title} {Active intracellular delivery of a {C}as9/sg{RNA} complex
  using ultrasound-propelled nanomotors},}\ }\href@noop {} {\bibfield
  {journal} {\bibinfo  {journal} {Angewandte Chemie International Edition}\
  }\textbf {\bibinfo {volume} {57}},\ \bibinfo {pages} {2657--2661} (\bibinfo
  {year} {2018})}\BibitemShut {NoStop}%
\bibitem [{\citenamefont {Sabrina}\ \emph {et~al.}(2018)\citenamefont
  {Sabrina}, \citenamefont {Tasinkevych}, \citenamefont {Ahmed}, \citenamefont
  {Brooks}, \citenamefont {{Olvera de la Cruz}}, \citenamefont {Mallouk},\ and\
  \citenamefont {Bishop}}]{SabrinaTABdlCMB2018}%
  \BibitemOpen
  \bibfield  {author} {\bibinfo {author} {\bibfnamefont {S.}~\bibnamefont
  {Sabrina}}, \bibinfo {author} {\bibfnamefont {M.}~\bibnamefont
  {Tasinkevych}}, \bibinfo {author} {\bibfnamefont {S.}~\bibnamefont {Ahmed}},
  \bibinfo {author} {\bibfnamefont {A.~M.}\ \bibnamefont {Brooks}}, \bibinfo
  {author} {\bibfnamefont {M.}~\bibnamefont {{Olvera de la Cruz}}}, \bibinfo
  {author} {\bibfnamefont {T.~E.}\ \bibnamefont {Mallouk}}, \ and\ \bibinfo
  {author} {\bibfnamefont {K.~J.~M.}\ \bibnamefont {Bishop}},\ }\bibfield
  {title} {\enquote {\bibinfo {title} {Shape-directed microspinners powered by
  ultrasound},}\ }\href@noop {} {\bibfield  {journal} {\bibinfo  {journal} {ACS
  Nano}\ }\textbf {\bibinfo {volume} {12}},\ \bibinfo {pages} {2939--2947}
  (\bibinfo {year} {2018})}\BibitemShut {NoStop}%
\bibitem [{\citenamefont {Ahmed}\ \emph
  {et~al.}(2016{\natexlab{b}})\citenamefont {Ahmed}, \citenamefont {Baasch},
  \citenamefont {Jang}, \citenamefont {Pane}, \citenamefont {Dual},\ and\
  \citenamefont {Nelson}}]{AhmedBJPDN2016}%
  \BibitemOpen
  \bibfield  {author} {\bibinfo {author} {\bibfnamefont {D.}~\bibnamefont
  {Ahmed}}, \bibinfo {author} {\bibfnamefont {T.}~\bibnamefont {Baasch}},
  \bibinfo {author} {\bibfnamefont {B.}~\bibnamefont {Jang}}, \bibinfo {author}
  {\bibfnamefont {S.}~\bibnamefont {Pane}}, \bibinfo {author} {\bibfnamefont
  {J.}~\bibnamefont {Dual}}, \ and\ \bibinfo {author} {\bibfnamefont {B.~J.}\
  \bibnamefont {Nelson}},\ }\bibfield  {title} {\enquote {\bibinfo {title}
  {Artificial swimmers propelled by acoustically activated flagella},}\
  }\href@noop {} {\bibfield  {journal} {\bibinfo  {journal} {Nano Letters}\
  }\textbf {\bibinfo {volume} {16}},\ \bibinfo {pages} {4968--4974} (\bibinfo
  {year} {2016}{\natexlab{b}})}\BibitemShut {NoStop}%
\bibitem [{\citenamefont {Zhou}\ \emph {et~al.}(2018)\citenamefont {Zhou},
  \citenamefont {Gao}, \citenamefont {Yang}, \citenamefont {Li}, \citenamefont
  {Shao}, \citenamefont {Zhang}, \citenamefont {Li},\ and\ \citenamefont
  {Li}}]{Zhou2018}%
  \BibitemOpen
  \bibfield  {author} {\bibinfo {author} {\bibfnamefont {D.}~\bibnamefont
  {Zhou}}, \bibinfo {author} {\bibfnamefont {Y.}~\bibnamefont {Gao}}, \bibinfo
  {author} {\bibfnamefont {J.}~\bibnamefont {Yang}}, \bibinfo {author}
  {\bibfnamefont {Y.~C.}\ \bibnamefont {Li}}, \bibinfo {author} {\bibfnamefont
  {G.}~\bibnamefont {Shao}}, \bibinfo {author} {\bibfnamefont {G.}~\bibnamefont
  {Zhang}}, \bibinfo {author} {\bibfnamefont {T.}~\bibnamefont {Li}}, \ and\
  \bibinfo {author} {\bibfnamefont {L.}~\bibnamefont {Li}},\ }\bibfield
  {title} {\enquote {\bibinfo {title} {Light-ultrasound driven collective
  {\textquotedblleft}firework{\textquotedblright} behavior of nanomotors},}\
  }\href@noop {} {\bibfield  {journal} {\bibinfo  {journal} {Advanced Science}\
  }\textbf {\bibinfo {volume} {5}},\ \bibinfo {pages} {1800122} (\bibinfo
  {year} {2018})}\BibitemShut {NoStop}%
\bibitem [{\citenamefont {Wang}\ \emph {et~al.}(2018)\citenamefont {Wang},
  \citenamefont {Gao}, \citenamefont {Wang}, \citenamefont {Sun}, \citenamefont
  {Guo}, \citenamefont {Xie},\ and\ \citenamefont {He}}]{WangGWSGXH2018}%
  \BibitemOpen
  \bibfield  {author} {\bibinfo {author} {\bibfnamefont {D.}~\bibnamefont
  {Wang}}, \bibinfo {author} {\bibfnamefont {C.}~\bibnamefont {Gao}}, \bibinfo
  {author} {\bibfnamefont {W.}~\bibnamefont {Wang}}, \bibinfo {author}
  {\bibfnamefont {M.}~\bibnamefont {Sun}}, \bibinfo {author} {\bibfnamefont
  {B.}~\bibnamefont {Guo}}, \bibinfo {author} {\bibfnamefont {H.}~\bibnamefont
  {Xie}}, \ and\ \bibinfo {author} {\bibfnamefont {Q.}~\bibnamefont {He}},\
  }\bibfield  {title} {\enquote {\bibinfo {title} {Shape-transformable, fusible
  rodlike swimming liquid metal nanomachine},}\ }\href@noop {} {\bibfield
  {journal} {\bibinfo  {journal} {ACS Nano}\ }\textbf {\bibinfo {volume}
  {12}},\ \bibinfo {pages} {10212--10220} (\bibinfo {year} {2018})}\BibitemShut
  {NoStop}%
\bibitem [{\citenamefont {{Esteban-Fern{\'a}ndez de {\'A}vila}}\ \emph
  {et~al.}(2018)\citenamefont {{Esteban-Fern{\'a}ndez de {\'A}vila}},
  \citenamefont {Angsantikul}, \citenamefont {{Ram{\'\i}rez-Herrera}},
  \citenamefont {Soto}, \citenamefont {Teymourian}, \citenamefont {Dehaini},
  \citenamefont {Chen}, \citenamefont {Zhang},\ and\ \citenamefont
  {Wang}}]{EstebanEtAl2018}%
  \BibitemOpen
  \bibfield  {author} {\bibinfo {author} {\bibfnamefont {B.}~\bibnamefont
  {{Esteban-Fern{\'a}ndez de {\'A}vila}}}, \bibinfo {author} {\bibfnamefont
  {P.}~\bibnamefont {Angsantikul}}, \bibinfo {author} {\bibfnamefont {D.~E.}\
  \bibnamefont {{Ram{\'\i}rez-Herrera}}}, \bibinfo {author} {\bibfnamefont
  {F.}~\bibnamefont {Soto}}, \bibinfo {author} {\bibfnamefont {H.}~\bibnamefont
  {Teymourian}}, \bibinfo {author} {\bibfnamefont {D.}~\bibnamefont {Dehaini}},
  \bibinfo {author} {\bibfnamefont {Y.}~\bibnamefont {Chen}}, \bibinfo {author}
  {\bibfnamefont {L.}~\bibnamefont {Zhang}}, \ and\ \bibinfo {author}
  {\bibfnamefont {J.}~\bibnamefont {Wang}},\ }\bibfield  {title} {\enquote
  {\bibinfo {title} {Hybrid biomembrane{\textendash}functionalized nanorobots
  for concurrent removal of pathogenic bacteria and toxins},}\ }\href@noop {}
  {\bibfield  {journal} {\bibinfo  {journal} {Science Robotics}\ }\textbf
  {\bibinfo {volume} {3}},\ \bibinfo {pages} {eaat0485} (\bibinfo {year}
  {2018})}\BibitemShut {NoStop}%
\bibitem [{\citenamefont {Lu}\ \emph {et~al.}(2019)\citenamefont {Lu},
  \citenamefont {Shen}, \citenamefont {Wang}, \citenamefont {Zhao},
  \citenamefont {Peng},\ and\ \citenamefont {Liu}}]{LuSZWPL2019}%
  \BibitemOpen
  \bibfield  {author} {\bibinfo {author} {\bibfnamefont {X.}~\bibnamefont
  {Lu}}, \bibinfo {author} {\bibfnamefont {H.}~\bibnamefont {Shen}}, \bibinfo
  {author} {\bibfnamefont {Z.}~\bibnamefont {Wang}}, \bibinfo {author}
  {\bibfnamefont {K.}~\bibnamefont {Zhao}}, \bibinfo {author} {\bibfnamefont
  {H.}~\bibnamefont {Peng}}, \ and\ \bibinfo {author} {\bibfnamefont
  {W.}~\bibnamefont {Liu}},\ }\bibfield  {title} {\enquote {\bibinfo {title}
  {{Micro/Nano} machines driven by ultrasound power sources},}\ }\href@noop {}
  {\bibfield  {journal} {\bibinfo  {journal} {Chemistry -- An Asian Journal}\
  }\textbf {\bibinfo {volume} {14}},\ \bibinfo {pages} {2406--2416} (\bibinfo
  {year} {2019})}\BibitemShut {NoStop}%
\bibitem [{\citenamefont {Tang}\ \emph {et~al.}(2019)\citenamefont {Tang},
  \citenamefont {Zhang}, \citenamefont {Zhao}, \citenamefont {Talaat},
  \citenamefont {Soto}, \citenamefont {Karshalev}, \citenamefont {Chen},
  \citenamefont {Hu}, \citenamefont {Lu}, \citenamefont {Li} \emph
  {et~al.}}]{TangEtAl2019}%
  \BibitemOpen
  \bibfield  {author} {\bibinfo {author} {\bibfnamefont {S.}~\bibnamefont
  {Tang}}, \bibinfo {author} {\bibfnamefont {F.}~\bibnamefont {Zhang}},
  \bibinfo {author} {\bibfnamefont {J.}~\bibnamefont {Zhao}}, \bibinfo {author}
  {\bibfnamefont {W.}~\bibnamefont {Talaat}}, \bibinfo {author} {\bibfnamefont
  {F.}~\bibnamefont {Soto}}, \bibinfo {author} {\bibfnamefont {E.}~\bibnamefont
  {Karshalev}}, \bibinfo {author} {\bibfnamefont {C.}~\bibnamefont {Chen}},
  \bibinfo {author} {\bibfnamefont {Z.}~\bibnamefont {Hu}}, \bibinfo {author}
  {\bibfnamefont {X.}~\bibnamefont {Lu}}, \bibinfo {author} {\bibfnamefont
  {J.}~\bibnamefont {Li}},  \emph {et~al.},\ }\bibfield  {title} {\enquote
  {\bibinfo {title} {Structure-dependent optical modulation of propulsion and
  collective behavior of acoustic/light-driven hybrid microbowls},}\
  }\href@noop {} {\bibfield  {journal} {\bibinfo  {journal} {Advanced
  Functional Materials}\ }\textbf {\bibinfo {volume} {29}},\ \bibinfo {pages}
  {1809003} (\bibinfo {year} {2019})}\BibitemShut {NoStop}%
\bibitem [{\citenamefont {Qualliotine}\ \emph {et~al.}(2019)\citenamefont
  {Qualliotine}, \citenamefont {Bolat}, \citenamefont
  {{Beltr{\'a}n-Gast{\'e}lum}}, \citenamefont {{Esteban-Fern{\'a}ndez de
  {\'A}vila}}, \citenamefont {Wang},\ and\ \citenamefont
  {Califano}}]{QualliotineEtAl2019}%
  \BibitemOpen
  \bibfield  {author} {\bibinfo {author} {\bibfnamefont {J.~R.}\ \bibnamefont
  {Qualliotine}}, \bibinfo {author} {\bibfnamefont {G.}~\bibnamefont {Bolat}},
  \bibinfo {author} {\bibfnamefont {M.}~\bibnamefont
  {{Beltr{\'a}n-Gast{\'e}lum}}}, \bibinfo {author} {\bibfnamefont
  {B.}~\bibnamefont {{Esteban-Fern{\'a}ndez de {\'A}vila}}}, \bibinfo {author}
  {\bibfnamefont {J.}~\bibnamefont {Wang}}, \ and\ \bibinfo {author}
  {\bibfnamefont {J.~A.}\ \bibnamefont {Califano}},\ }\bibfield  {title}
  {\enquote {\bibinfo {title} {Acoustic nanomotors for detection of human
  papillomavirus-associated head and neck cancer},}\ }\href@noop {} {\bibfield
  {journal} {\bibinfo  {journal} {Otolaryngology--Head and Neck Surgery}\
  }\textbf {\bibinfo {volume} {161}},\ \bibinfo {pages} {814--822} (\bibinfo
  {year} {2019})}\BibitemShut {NoStop}%
\bibitem [{\citenamefont {Gao}\ \emph {et~al.}(2019)\citenamefont {Gao},
  \citenamefont {Lin}, \citenamefont {Wang}, \citenamefont {Wu}, \citenamefont
  {Xie},\ and\ \citenamefont {He}}]{GaoLWWXH2019}%
  \BibitemOpen
  \bibfield  {author} {\bibinfo {author} {\bibfnamefont {C.}~\bibnamefont
  {Gao}}, \bibinfo {author} {\bibfnamefont {Z.}~\bibnamefont {Lin}}, \bibinfo
  {author} {\bibfnamefont {D.}~\bibnamefont {Wang}}, \bibinfo {author}
  {\bibfnamefont {Z.}~\bibnamefont {Wu}}, \bibinfo {author} {\bibfnamefont
  {H.}~\bibnamefont {Xie}}, \ and\ \bibinfo {author} {\bibfnamefont
  {Q.}~\bibnamefont {He}},\ }\bibfield  {title} {\enquote {\bibinfo {title}
  {Red blood cell-mimicking micromotor for active photodynamic cancer
  therapy},}\ }\href@noop {} {\bibfield  {journal} {\bibinfo  {journal} {ACS
  Applied Materials \& Interfaces}\ }\textbf {\bibinfo {volume} {11}},\
  \bibinfo {pages} {23392--23400} (\bibinfo {year} {2019})}\BibitemShut
  {NoStop}%
\bibitem [{\citenamefont {Ren}\ \emph {et~al.}(2019)\citenamefont {Ren},
  \citenamefont {Nama}, \citenamefont {{McNeill}}, \citenamefont {Soto},
  \citenamefont {Yan}, \citenamefont {Liu}, \citenamefont {Wang}, \citenamefont
  {Wang},\ and\ \citenamefont {Mallouk}}]{RenEtAl2019}%
  \BibitemOpen
  \bibfield  {author} {\bibinfo {author} {\bibfnamefont {L.}~\bibnamefont
  {Ren}}, \bibinfo {author} {\bibfnamefont {N.}~\bibnamefont {Nama}}, \bibinfo
  {author} {\bibfnamefont {J.~M.}\ \bibnamefont {{McNeill}}}, \bibinfo {author}
  {\bibfnamefont {F.}~\bibnamefont {Soto}}, \bibinfo {author} {\bibfnamefont
  {Z.}~\bibnamefont {Yan}}, \bibinfo {author} {\bibfnamefont {W.}~\bibnamefont
  {Liu}}, \bibinfo {author} {\bibfnamefont {W.}~\bibnamefont {Wang}}, \bibinfo
  {author} {\bibfnamefont {J.}~\bibnamefont {Wang}}, \ and\ \bibinfo {author}
  {\bibfnamefont {T.~E.}\ \bibnamefont {Mallouk}},\ }\bibfield  {title}
  {\enquote {\bibinfo {title} {3{D} steerable, acoustically powered
  microswimmers for single-particle manipulation},}\ }\href@noop {} {\bibfield
  {journal} {\bibinfo  {journal} {Science Advances}\ }\textbf {\bibinfo
  {volume} {5}},\ \bibinfo {pages} {eaax3084} (\bibinfo {year}
  {2019})}\BibitemShut {NoStop}%
\bibitem [{\citenamefont {Bhuyan}\ \emph {et~al.}(2019)\citenamefont {Bhuyan},
  \citenamefont {Dutta}, \citenamefont {Bhattacharjee}, \citenamefont {Singh},
  \citenamefont {Ghosh},\ and\ \citenamefont
  {Bandyopadhyay}}]{BhuyanDBSGB2019}%
  \BibitemOpen
  \bibfield  {author} {\bibinfo {author} {\bibfnamefont {T.}~\bibnamefont
  {Bhuyan}}, \bibinfo {author} {\bibfnamefont {D.}~\bibnamefont {Dutta}},
  \bibinfo {author} {\bibfnamefont {M.}~\bibnamefont {Bhattacharjee}}, \bibinfo
  {author} {\bibfnamefont {A.}~\bibnamefont {Singh}}, \bibinfo {author}
  {\bibfnamefont {S.}~\bibnamefont {Ghosh}}, \ and\ \bibinfo {author}
  {\bibfnamefont {D.}~\bibnamefont {Bandyopadhyay}},\ }\bibfield  {title}
  {\enquote {\bibinfo {title} {Acoustic propulsion of vitamin {C} loaded
  teabots for targeted oxidative stress and amyloid therapeutics},}\
  }\href@noop {} {\bibfield  {journal} {\bibinfo  {journal} {ACS Applied Bio
  Materials}\ }\textbf {\bibinfo {volume} {2}},\ \bibinfo {pages} {4571--4582}
  (\bibinfo {year} {2019})}\BibitemShut {NoStop}%
\bibitem [{\citenamefont {Aghakhani}\ \emph {et~al.}(2020)\citenamefont
  {Aghakhani}, \citenamefont {Yasa}, \citenamefont {Wrede},\ and\ \citenamefont
  {Sitti}}]{AghakhaniYWS2020}%
  \BibitemOpen
  \bibfield  {author} {\bibinfo {author} {\bibfnamefont {A.}~\bibnamefont
  {Aghakhani}}, \bibinfo {author} {\bibfnamefont {O.}~\bibnamefont {Yasa}},
  \bibinfo {author} {\bibfnamefont {P.}~\bibnamefont {Wrede}}, \ and\ \bibinfo
  {author} {\bibfnamefont {M.}~\bibnamefont {Sitti}},\ }\bibfield  {title}
  {\enquote {\bibinfo {title} {Acoustically powered surface-slipping mobile
  microrobots},}\ }\href@noop {} {\bibfield  {journal} {\bibinfo  {journal}
  {Proceedings of the National Academy of Sciences U.S.A.}\ }\textbf {\bibinfo
  {volume} {117}},\ \bibinfo {pages} {3469--3477} (\bibinfo {year}
  {2020})}\BibitemShut {NoStop}%
\bibitem [{\citenamefont {Liu}\ and\ \citenamefont {Ruan}(2020)}]{LiuR2020}%
  \BibitemOpen
  \bibfield  {author} {\bibinfo {author} {\bibfnamefont {J.}~\bibnamefont
  {Liu}}\ and\ \bibinfo {author} {\bibfnamefont {H.}~\bibnamefont {Ruan}},\
  }\bibfield  {title} {\enquote {\bibinfo {title} {Modeling of an acoustically
  actuated artificial micro-swimmer},}\ }\href@noop {} {\bibfield  {journal}
  {\bibinfo  {journal} {Bioinspiration \& Biomimetics}\ }\textbf {\bibinfo
  {volume} {15}},\ \bibinfo {pages} {036002} (\bibinfo {year}
  {2020})}\BibitemShut {NoStop}%
\bibitem [{\citenamefont {Vo{\ss}}\ and\ \citenamefont
  {Wittkowski}(2020)}]{VossW2020}%
  \BibitemOpen
  \bibfield  {author} {\bibinfo {author} {\bibfnamefont {J.}~\bibnamefont
  {Vo{\ss}}}\ and\ \bibinfo {author} {\bibfnamefont {R.}~\bibnamefont
  {Wittkowski}},\ }\bibfield  {title} {\enquote {\bibinfo {title} {On the
  shape-dependent propulsion of nano- and microparticles by traveling
  ultrasound waves},}\ }\href@noop {} {\bibfield  {journal} {\bibinfo
  {journal} {Nanoscale Advances}\ }\textbf {\bibinfo {volume} {2}},\ \bibinfo
  {pages} {3890--3899} (\bibinfo {year} {2020})}\BibitemShut {NoStop}%
\bibitem [{\citenamefont {Valdez-Gardu{\~n}o}\ \emph
  {et~al.}(2020)\citenamefont {Valdez-Gardu{\~n}o}, \citenamefont
  {Leal-Estrada}, \citenamefont {Oliveros-Mata}, \citenamefont
  {Sandoval-Bojorquez}, \citenamefont {Soto}, \citenamefont {Wang},\ and\
  \citenamefont {Garcia-Gradilla}}]{ValdezLOESSWG2020}%
  \BibitemOpen
  \bibfield  {author} {\bibinfo {author} {\bibfnamefont {M.}~\bibnamefont
  {Valdez-Gardu{\~n}o}}, \bibinfo {author} {\bibfnamefont {M.}~\bibnamefont
  {Leal-Estrada}}, \bibinfo {author} {\bibfnamefont {E.~S.}\ \bibnamefont
  {Oliveros-Mata}}, \bibinfo {author} {\bibfnamefont {D.~I.}\ \bibnamefont
  {Sandoval-Bojorquez}}, \bibinfo {author} {\bibfnamefont {F.}~\bibnamefont
  {Soto}}, \bibinfo {author} {\bibfnamefont {J.}~\bibnamefont {Wang}}, \ and\
  \bibinfo {author} {\bibfnamefont {V.}~\bibnamefont {Garcia-Gradilla}},\
  }\bibfield  {title} {\enquote {\bibinfo {title} {Density asymmetry driven
  propulsion of ultrasound-powered {J}anus micromotors},}\ }\href@noop {}
  {\bibfield  {journal} {\bibinfo  {journal} {Advanced Functional Materials}\
  }\textbf {\bibinfo {volume} {30}},\ \bibinfo {pages} {2004043} (\bibinfo
  {year} {2020})}\BibitemShut {NoStop}%
\bibitem [{\citenamefont {Dumy}\ \emph {et~al.}(2020)\citenamefont {Dumy},
  \citenamefont {{Jeger-Madiot}}, \citenamefont {{Benoit-Gonin}}, \citenamefont
  {Mallouk}, \citenamefont {Hoyos},\ and\ \citenamefont
  {Aider}}]{DumyJMBGMHA2020}%
  \BibitemOpen
  \bibfield  {author} {\bibinfo {author} {\bibfnamefont {G.}~\bibnamefont
  {Dumy}}, \bibinfo {author} {\bibfnamefont {N.}~\bibnamefont
  {{Jeger-Madiot}}}, \bibinfo {author} {\bibfnamefont {X.}~\bibnamefont
  {{Benoit-Gonin}}}, \bibinfo {author} {\bibfnamefont {T.}~\bibnamefont
  {Mallouk}}, \bibinfo {author} {\bibfnamefont {M.}~\bibnamefont {Hoyos}}, \
  and\ \bibinfo {author} {\bibfnamefont {J.}~\bibnamefont {Aider}},\ }\bibfield
   {title} {\enquote {\bibinfo {title} {Acoustic manipulation of dense nanorods
  in microgravity},}\ }\href@noop {} {\bibfield  {journal} {\bibinfo  {journal}
  {Microgravity Science and Technology}\ }\textbf {\bibinfo {volume} {32}},\
  \bibinfo {pages} {1159--1174} (\bibinfo {year} {2020})}\BibitemShut {NoStop}%
\bibitem [{\citenamefont {McNeill}\ \emph {et~al.}(2020)\citenamefont
  {McNeill}, \citenamefont {Nama}, \citenamefont {Braxton},\ and\ \citenamefont
  {Mallouk}}]{McneillNBM2020}%
  \BibitemOpen
  \bibfield  {author} {\bibinfo {author} {\bibfnamefont {J.}~\bibnamefont
  {McNeill}}, \bibinfo {author} {\bibfnamefont {N.}~\bibnamefont {Nama}},
  \bibinfo {author} {\bibfnamefont {J.}~\bibnamefont {Braxton}}, \ and\
  \bibinfo {author} {\bibfnamefont {T.}~\bibnamefont {Mallouk}},\ }\bibfield
  {title} {\enquote {\bibinfo {title} {Wafer-scale fabrication of micro- to
  nanoscale bubble swimmers and their fast autonomous propulsion by
  ultrasound},}\ }\href@noop {} {\bibfield  {journal} {\bibinfo  {journal} {ACS
  Nano}\ }\textbf {\bibinfo {volume} {14}},\ \bibinfo {pages} {7520--7528}
  (\bibinfo {year} {2020})}\BibitemShut {NoStop}%
\bibitem [{\citenamefont {McNeill}\ \emph {et~al.}(2021)\citenamefont
  {McNeill}, \citenamefont {Sinai}, \citenamefont {Wang}, \citenamefont
  {Oliver}, \citenamefont {Lauga}, \citenamefont {Nadal},\ and\ \citenamefont
  {Mallouk}}]{McneillSWOLNM2021}%
  \BibitemOpen
  \bibfield  {author} {\bibinfo {author} {\bibfnamefont {J.}~\bibnamefont
  {McNeill}}, \bibinfo {author} {\bibfnamefont {N.}~\bibnamefont {Sinai}},
  \bibinfo {author} {\bibfnamefont {J.}~\bibnamefont {Wang}}, \bibinfo {author}
  {\bibfnamefont {V.}~\bibnamefont {Oliver}}, \bibinfo {author} {\bibfnamefont
  {E.}~\bibnamefont {Lauga}}, \bibinfo {author} {\bibfnamefont
  {F.}~\bibnamefont {Nadal}}, \ and\ \bibinfo {author} {\bibfnamefont
  {T.}~\bibnamefont {Mallouk}},\ }\bibfield  {title} {\enquote {\bibinfo
  {title} {Purely viscous acoustic propulsion of bimetallic rods},}\
  }\href@noop {} {\bibfield  {journal} {\bibinfo  {journal} {Physical Review
  Fluids}\ }\textbf {\bibinfo {volume} {6}},\ \bibinfo {pages} {L092201}
  (\bibinfo {year} {2021})}\BibitemShut {NoStop}%
\bibitem [{\citenamefont {Vo{\ss}}\ and\ \citenamefont
  {Wittkowski}(2022{\natexlab{a}})}]{VossW2021}%
  \BibitemOpen
  \bibfield  {author} {\bibinfo {author} {\bibfnamefont {J.}~\bibnamefont
  {Vo{\ss}}}\ and\ \bibinfo {author} {\bibfnamefont {R.}~\bibnamefont
  {Wittkowski}},\ }\bibfield  {title} {\enquote {\bibinfo {title} {Acoustically
  propelled nano- and microcones: fast forward and backward motion},}\
  }\href@noop {} {\bibfield  {journal} {\bibinfo  {journal} {Nanoscale
  Advances}\ }\textbf {\bibinfo {volume} {4}},\ \bibinfo {pages} {281--293}
  (\bibinfo {year} {2022}{\natexlab{a}})}\BibitemShut {NoStop}%
\bibitem [{\citenamefont {Mohanty}\ \emph {et~al.}(2021)\citenamefont
  {Mohanty}, \citenamefont {Zhang}, \citenamefont {McNeill}, \citenamefont
  {Kuenen}, \citenamefont {Linde}, \citenamefont {Rouwkema},\ and\
  \citenamefont {Misra}}]{MohantyEtAl2021}%
  \BibitemOpen
  \bibfield  {author} {\bibinfo {author} {\bibfnamefont {S.}~\bibnamefont
  {Mohanty}}, \bibinfo {author} {\bibfnamefont {J.}~\bibnamefont {Zhang}},
  \bibinfo {author} {\bibfnamefont {J.}~\bibnamefont {McNeill}}, \bibinfo
  {author} {\bibfnamefont {T.}~\bibnamefont {Kuenen}}, \bibinfo {author}
  {\bibfnamefont {F.}~\bibnamefont {Linde}}, \bibinfo {author} {\bibfnamefont
  {J.}~\bibnamefont {Rouwkema}}, \ and\ \bibinfo {author} {\bibfnamefont
  {S.}~\bibnamefont {Misra}},\ }\bibfield  {title} {\enquote {\bibinfo {title}
  {Acoustically-actuated bubble-powered rotational micro-propellers},}\
  }\href@noop {} {\bibfield  {journal} {\bibinfo  {journal} {Sensors and
  Actuators B: Chemical}\ }\textbf {\bibinfo {volume} {347}},\ \bibinfo {pages}
  {130589} (\bibinfo {year} {2021})}\BibitemShut {NoStop}%
\bibitem [{\citenamefont {Li}\ \emph {et~al.}(2022)\citenamefont {Li},
  \citenamefont {{Mayorga-Martinez}}, \citenamefont {Ohl},\ and\ \citenamefont
  {Pumera}}]{LiMMOP2021}%
  \BibitemOpen
  \bibfield  {author} {\bibinfo {author} {\bibfnamefont {J.}~\bibnamefont
  {Li}}, \bibinfo {author} {\bibfnamefont {C.}~\bibnamefont
  {{Mayorga-Martinez}}}, \bibinfo {author} {\bibfnamefont {C.}~\bibnamefont
  {Ohl}}, \ and\ \bibinfo {author} {\bibfnamefont {M.}~\bibnamefont {Pumera}},\
  }\bibfield  {title} {\enquote {\bibinfo {title} {Ultrasonically propelled
  micro- and nanorobots},}\ }\href@noop {} {\bibfield  {journal} {\bibinfo
  {journal} {Advanced Functional Materials}\ }\textbf {\bibinfo {volume}
  {32}},\ \bibinfo {pages} {2102265} (\bibinfo {year} {2022})}\BibitemShut
  {NoStop}%
\bibitem [{\citenamefont {Vo{\ss}}\ and\ \citenamefont
  {Wittkowski}(2022{\natexlab{b}})}]{VossW2022orientation}%
  \BibitemOpen
  \bibfield  {author} {\bibinfo {author} {\bibfnamefont {J.}~\bibnamefont
  {Vo{\ss}}}\ and\ \bibinfo {author} {\bibfnamefont {R.}~\bibnamefont
  {Wittkowski}},\ }\bibfield  {title} {\enquote {\bibinfo {title}
  {Orientation-dependent propulsion of triangular nano- and microparticles by a
  traveling ultrasound wave},}\ }\href@noop {} {\bibfield  {journal} {\bibinfo
  {journal} {ACS Nano}\ ,\ \bibinfo {pages} {provisionally accepted}} (\bibinfo
  {year} {2022}{\natexlab{b}})}\BibitemShut {NoStop}%
\bibitem [{\citenamefont {Vo{\ss}}\ and\ \citenamefont
  {Wittkowski}(2022{\natexlab{c}})}]{VossW2022acoustica}%
  \BibitemOpen
  \bibfield  {author} {\bibinfo {author} {\bibfnamefont {J.}~\bibnamefont
  {Vo{\ss}}}\ and\ \bibinfo {author} {\bibfnamefont {R.}~\bibnamefont
  {Wittkowski}},\ }\bibfield  {title} {\enquote {\bibinfo {title} {Acoustic
  propulsion of nano- and microcones: dependence on particle size, acoustic
  energy density, and sound frequency},}\ }\href@noop {} {\bibfield  {journal}
  {\bibinfo  {journal} {submitted}\ } (\bibinfo {year}
  {2022}{\natexlab{c}})}\BibitemShut {NoStop}%
\bibitem [{\citenamefont {Vo{\ss}}\ and\ \citenamefont
  {Wittkowski}(2022{\natexlab{d}})}]{VossW2022acousticb}%
  \BibitemOpen
  \bibfield  {author} {\bibinfo {author} {\bibfnamefont {J.}~\bibnamefont
  {Vo{\ss}}}\ and\ \bibinfo {author} {\bibfnamefont {R.}~\bibnamefont
  {Wittkowski}},\ }\bibfield  {title} {\enquote {\bibinfo {title} {Acoustic
  propulsion of nano- and microcones: dependence on the viscosity of the
  surrounding fluid},}\ }\href@noop {} {\bibfield  {journal} {\bibinfo
  {journal} {submitted}\ } (\bibinfo {year} {2022}{\natexlab{d}})}\BibitemShut
  {NoStop}%
\bibitem [{\citenamefont {Vo{\ss}}\ and\ \citenamefont
  {Wittkowski}(2022{\natexlab{e}})}]{VossW2022microspinner}%
  \BibitemOpen
  \bibfield  {author} {\bibinfo {author} {\bibfnamefont {J.}~\bibnamefont
  {Vo{\ss}}}\ and\ \bibinfo {author} {\bibfnamefont {R.}~\bibnamefont
  {Wittkowski}},\ }\bibfield  {title} {\enquote {\bibinfo {title}
  {Ultrasound-propelled nano- and microspinners},}\ }\href@noop {} {\bibfield
  {journal} {\bibinfo  {journal} {submitted}\ } (\bibinfo {year}
  {2022}{\natexlab{e}})}\BibitemShut {NoStop}%
\bibitem [{\citenamefont {Li}\ \emph {et~al.}(2017)\citenamefont {Li},
  \citenamefont {{Esteban-Fern{\'a}ndez de {\'A}vila}}, \citenamefont {Gao},
  \citenamefont {Zhang},\ and\ \citenamefont {Wang}}]{LiEFdAGZW2017}%
  \BibitemOpen
  \bibfield  {author} {\bibinfo {author} {\bibfnamefont {J.}~\bibnamefont
  {Li}}, \bibinfo {author} {\bibfnamefont {B.}~\bibnamefont
  {{Esteban-Fern{\'a}ndez de {\'A}vila}}}, \bibinfo {author} {\bibfnamefont
  {W.}~\bibnamefont {Gao}}, \bibinfo {author} {\bibfnamefont {L.}~\bibnamefont
  {Zhang}}, \ and\ \bibinfo {author} {\bibfnamefont {J.}~\bibnamefont {Wang}},\
  }\bibfield  {title} {\enquote {\bibinfo {title} {Micro/{N}anorobots for
  biomedicine: delivery, surgery, sensing, and detoxification},}\ }\href@noop
  {} {\bibfield  {journal} {\bibinfo  {journal} {Science Robotics}\ }\textbf
  {\bibinfo {volume} {2}},\ \bibinfo {pages} {eaam6431} (\bibinfo {year}
  {2017})}\BibitemShut {NoStop}%
\bibitem [{\citenamefont {Peng}\ \emph {et~al.}(2017)\citenamefont {Peng},
  \citenamefont {Tu},\ and\ \citenamefont {Wilson}}]{PengTW2017}%
  \BibitemOpen
  \bibfield  {author} {\bibinfo {author} {\bibfnamefont {F.}~\bibnamefont
  {Peng}}, \bibinfo {author} {\bibfnamefont {Y.}~\bibnamefont {Tu}}, \ and\
  \bibinfo {author} {\bibfnamefont {D.~A.}\ \bibnamefont {Wilson}},\ }\bibfield
   {title} {\enquote {\bibinfo {title} {Micro/{N}anomotors towards in vivo
  application: cell, tissue and biofluid},}\ }\href@noop {} {\bibfield
  {journal} {\bibinfo  {journal} {Chemical Society Reviews}\ }\textbf {\bibinfo
  {volume} {46}},\ \bibinfo {pages} {5289--5310} (\bibinfo {year}
  {2017})}\BibitemShut {NoStop}%
\bibitem [{\citenamefont {Soto}\ and\ \citenamefont
  {Chrostowski}(2018)}]{SotoC2018}%
  \BibitemOpen
  \bibfield  {author} {\bibinfo {author} {\bibfnamefont {F.}~\bibnamefont
  {Soto}}\ and\ \bibinfo {author} {\bibfnamefont {R.}~\bibnamefont
  {Chrostowski}},\ }\bibfield  {title} {\enquote {\bibinfo {title} {Frontiers
  of medical micro/nanorobotics: in vivo applications and commercialization
  perspectives toward clinical uses},}\ }\href@noop {} {\bibfield  {journal}
  {\bibinfo  {journal} {Frontiers in Bioengineering and Biotechnology}\
  }\textbf {\bibinfo {volume} {6}},\ \bibinfo {pages} {170} (\bibinfo {year}
  {2018})}\BibitemShut {NoStop}%
\bibitem [{\citenamefont {Wang}\ \emph {et~al.}(2020)\citenamefont {Wang},
  \citenamefont {Gao}, \citenamefont {Zhou}, \citenamefont {Lin},\ and\
  \citenamefont {He}}]{WangGZLH2020}%
  \BibitemOpen
  \bibfield  {author} {\bibinfo {author} {\bibfnamefont {D.}~\bibnamefont
  {Wang}}, \bibinfo {author} {\bibfnamefont {C.}~\bibnamefont {Gao}}, \bibinfo
  {author} {\bibfnamefont {C.}~\bibnamefont {Zhou}}, \bibinfo {author}
  {\bibfnamefont {Z.}~\bibnamefont {Lin}}, \ and\ \bibinfo {author}
  {\bibfnamefont {Q.}~\bibnamefont {He}},\ }\bibfield  {title} {\enquote
  {\bibinfo {title} {Leukocyte membrane-coated liquid metal nanoswimmers for
  actively targeted delivery and synergistic chemophotothermal therapy},}\
  }\href@noop {} {\bibfield  {journal} {\bibinfo  {journal} {Research}\
  }\textbf {\bibinfo {volume} {2020}},\ \bibinfo {pages} {3676954} (\bibinfo
  {year} {2020})}\BibitemShut {NoStop}%
\bibitem [{\citenamefont {Wang}\ and\ \citenamefont {Zhou}(2021)}]{WangZ2021}%
  \BibitemOpen
  \bibfield  {author} {\bibinfo {author} {\bibfnamefont {W.}~\bibnamefont
  {Wang}}\ and\ \bibinfo {author} {\bibfnamefont {C.}~\bibnamefont {Zhou}},\
  }\bibfield  {title} {\enquote {\bibinfo {title} {A journey of nanomotors for
  targeted cancer therapy: Principles, challenges, and a critical review of the
  state-of-the-art},}\ }\href@noop {} {\bibfield  {journal} {\bibinfo
  {journal} {Advanced Healthcare Materials}\ }\textbf {\bibinfo {volume}
  {10}},\ \bibinfo {pages} {2001236} (\bibinfo {year} {2021})}\BibitemShut
  {NoStop}%
\bibitem [{\citenamefont {{Leal-Estrada}}\ \emph {et~al.}(2021)\citenamefont
  {{Leal-Estrada}}, \citenamefont {{Valdez-Gardu{\~n}o}}, \citenamefont
  {Soto},\ and\ \citenamefont {{Garcia-Gradilla}}}]{Leal2021}%
  \BibitemOpen
  \bibfield  {author} {\bibinfo {author} {\bibfnamefont {M.}~\bibnamefont
  {{Leal-Estrada}}}, \bibinfo {author} {\bibfnamefont {M.}~\bibnamefont
  {{Valdez-Gardu{\~n}o}}}, \bibinfo {author} {\bibfnamefont {F.}~\bibnamefont
  {Soto}}, \ and\ \bibinfo {author} {\bibfnamefont {V.}~\bibnamefont
  {{Garcia-Gradilla}}},\ }\bibfield  {title} {\enquote {\bibinfo {title}
  {Engineering ultrasound fields to power medical micro/nanorobots},}\
  }\href@noop {} {\bibfield  {journal} {\bibinfo  {journal} {Current Robotics
  Reports}\ }\textbf {\bibinfo {volume} {2}},\ \bibinfo {pages} {21--32}
  (\bibinfo {year} {2021})}\BibitemShut {NoStop}%
\bibitem [{\citenamefont {Nitschke}\ and\ \citenamefont
  {Wittkowski}(2021)}]{NitschkeW2021}%
  \BibitemOpen
  \bibfield  {author} {\bibinfo {author} {\bibfnamefont {T.}~\bibnamefont
  {Nitschke}}\ and\ \bibinfo {author} {\bibfnamefont {R.}~\bibnamefont
  {Wittkowski}},\ }\bibfield  {title} {\enquote {\bibinfo {title} {Collective
  guiding of acoustically propelled nano- and microparticles for medical
  applications},}\ }\href@noop {} {\bibfield  {journal} {\bibinfo  {journal}
  {arXiv:2112.13676}\ } (\bibinfo {year} {2021})}\BibitemShut {NoStop}%
\bibitem [{\citenamefont {Jun}\ and\ \citenamefont {Hess}(2010)}]{JunH2010}%
  \BibitemOpen
  \bibfield  {author} {\bibinfo {author} {\bibfnamefont {I.}~\bibnamefont
  {Jun}}\ and\ \bibinfo {author} {\bibfnamefont {H.}~\bibnamefont {Hess}},\
  }\bibfield  {title} {\enquote {\bibinfo {title} {A biomimetic, self-pumping
  membrane},}\ }\href@noop {} {\bibfield  {journal} {\bibinfo  {journal}
  {Advanced Materials}\ }\textbf {\bibinfo {volume} {22}},\ \bibinfo {pages}
  {4823--4825} (\bibinfo {year} {2010})}\BibitemShut {NoStop}%
\bibitem [{\citenamefont {McDermott}\ \emph {et~al.}(2012)\citenamefont
  {McDermott}, \citenamefont {Kar}, \citenamefont {Daher}, \citenamefont
  {Klara}, \citenamefont {Wang}, \citenamefont {Sen},\ and\ \citenamefont
  {Velegol}}]{McdermottKDKWSV2012}%
  \BibitemOpen
  \bibfield  {author} {\bibinfo {author} {\bibfnamefont {J.}~\bibnamefont
  {McDermott}}, \bibinfo {author} {\bibfnamefont {A.}~\bibnamefont {Kar}},
  \bibinfo {author} {\bibfnamefont {M.}~\bibnamefont {Daher}}, \bibinfo
  {author} {\bibfnamefont {S.}~\bibnamefont {Klara}}, \bibinfo {author}
  {\bibfnamefont {G.}~\bibnamefont {Wang}}, \bibinfo {author} {\bibfnamefont
  {A.}~\bibnamefont {Sen}}, \ and\ \bibinfo {author} {\bibfnamefont
  {D.}~\bibnamefont {Velegol}},\ }\bibfield  {title} {\enquote {\bibinfo
  {title} {Self-generated diffusioosmotic flows from calcium carbonate
  micropumps},}\ }\href@noop {} {\bibfield  {journal} {\bibinfo  {journal}
  {Langmuir}\ }\textbf {\bibinfo {volume} {28}},\ \bibinfo {pages}
  {15491--15497} (\bibinfo {year} {2012})}\BibitemShut {NoStop}%
\bibitem [{\citenamefont {Fratzl}\ \emph {et~al.}(2021)\citenamefont {Fratzl},
  \citenamefont {Friedman}, \citenamefont {Krauthausen},\ and\ \citenamefont
  {Sch{\"a}ffner}}]{FratzlFKS2021}%
  \BibitemOpen
  \bibinfo {editor} {\bibfnamefont {P.}~\bibnamefont {Fratzl}}, \bibinfo
  {editor} {\bibfnamefont {M.}~\bibnamefont {Friedman}}, \bibinfo {editor}
  {\bibfnamefont {K.}~\bibnamefont {Krauthausen}}, \ and\ \bibinfo {editor}
  {\bibfnamefont {W.}~\bibnamefont {Sch{\"a}ffner}},\ eds.,\ \href@noop {}
  {\emph {\bibinfo {title} {Active Materials}}},\ \bibinfo {edition} {1st}\
  ed.\ (\bibinfo  {publisher} {De Gruyter},\ \bibinfo {address} {Berlin},\
  \bibinfo {year} {2021})\BibitemShut {NoStop}%
\bibitem [{\citenamefont {Li}\ \emph {et~al.}(2015)\citenamefont {Li},
  \citenamefont {Li}, \citenamefont {Xu}, \citenamefont {Kiristi},
  \citenamefont {Liu}, \citenamefont {Wu},\ and\ \citenamefont
  {Wang}}]{LILXKLWW2015}%
  \BibitemOpen
  \bibfield  {author} {\bibinfo {author} {\bibfnamefont {J.}~\bibnamefont
  {Li}}, \bibinfo {author} {\bibfnamefont {T.}~\bibnamefont {Li}}, \bibinfo
  {author} {\bibfnamefont {T.}~\bibnamefont {Xu}}, \bibinfo {author}
  {\bibfnamefont {M.}~\bibnamefont {Kiristi}}, \bibinfo {author} {\bibfnamefont
  {W.}~\bibnamefont {Liu}}, \bibinfo {author} {\bibfnamefont {Z.}~\bibnamefont
  {Wu}}, \ and\ \bibinfo {author} {\bibfnamefont {J.}~\bibnamefont {Wang}},\
  }\bibfield  {title} {\enquote {\bibinfo {title} {Magneto-acoustic hybrid
  nanomotor},}\ }\href@noop {} {\bibfield  {journal} {\bibinfo  {journal} {Nano
  Letters}\ }\textbf {\bibinfo {volume} {15}},\ \bibinfo {pages} {4814--4821}
  (\bibinfo {year} {2015})}\BibitemShut {NoStop}%
\bibitem [{\citenamefont {Ren}\ \emph {et~al.}(2018)\citenamefont {Ren},
  \citenamefont {Wang},\ and\ \citenamefont {Mallouk}}]{RenWM2018}%
  \BibitemOpen
  \bibfield  {author} {\bibinfo {author} {\bibfnamefont {L.}~\bibnamefont
  {Ren}}, \bibinfo {author} {\bibfnamefont {W.}~\bibnamefont {Wang}}, \ and\
  \bibinfo {author} {\bibfnamefont {T.~E.}\ \bibnamefont {Mallouk}},\
  }\bibfield  {title} {\enquote {\bibinfo {title} {Two forces are better than
  one: combining chemical and acoustic propulsion for enhanced micromotor
  functionality},}\ }\href@noop {} {\bibfield  {journal} {\bibinfo  {journal}
  {Accounts of Chemical Research}\ }\textbf {\bibinfo {volume} {51}},\ \bibinfo
  {pages} {1948--1956} (\bibinfo {year} {2018})}\BibitemShut {NoStop}%
\bibitem [{\citenamefont {Xu}\ \emph {et~al.}(2015)\citenamefont {Xu},
  \citenamefont {Soto}, \citenamefont {Gao}, \citenamefont {Dong},
  \citenamefont {{Garcia-Gradilla}}, \citenamefont {Magana}, \citenamefont
  {Zhang},\ and\ \citenamefont {Wang}}]{XuEtAl2015}%
  \BibitemOpen
  \bibfield  {author} {\bibinfo {author} {\bibfnamefont {T.}~\bibnamefont
  {Xu}}, \bibinfo {author} {\bibfnamefont {F.}~\bibnamefont {Soto}}, \bibinfo
  {author} {\bibfnamefont {W.}~\bibnamefont {Gao}}, \bibinfo {author}
  {\bibfnamefont {R.}~\bibnamefont {Dong}}, \bibinfo {author} {\bibfnamefont
  {V.}~\bibnamefont {{Garcia-Gradilla}}}, \bibinfo {author} {\bibfnamefont
  {E.}~\bibnamefont {Magana}}, \bibinfo {author} {\bibfnamefont
  {X.}~\bibnamefont {Zhang}}, \ and\ \bibinfo {author} {\bibfnamefont
  {J.}~\bibnamefont {Wang}},\ }\bibfield  {title} {\enquote {\bibinfo {title}
  {Reversible swarming and separation of self-propelled chemically powered
  nanomotors under acoustic fields},}\ }\href@noop {} {\bibfield  {journal}
  {\bibinfo  {journal} {Journal of the American Chemical Society}\ }\textbf
  {\bibinfo {volume} {137}},\ \bibinfo {pages} {2163--2166} (\bibinfo {year}
  {2015})}\BibitemShut {NoStop}%
\bibitem [{\citenamefont {{Beltr{\'a}n-Gast{\'e}lum}}\ \emph
  {et~al.}(2019)\citenamefont {{Beltr{\'a}n-Gast{\'e}lum}}, \citenamefont
  {{Esteban-Fern{\'a}ndez de {\'A}vila}}, \citenamefont {Gong}, \citenamefont
  {Venugopalan}, \citenamefont {Hianik}, \citenamefont {Wang},\ and\
  \citenamefont {Subjakova}}]{BeltranEtAl2019}%
  \BibitemOpen
  \bibfield  {author} {\bibinfo {author} {\bibfnamefont {M.}~\bibnamefont
  {{Beltr{\'a}n-Gast{\'e}lum}}}, \bibinfo {author} {\bibfnamefont
  {B.}~\bibnamefont {{Esteban-Fern{\'a}ndez de {\'A}vila}}}, \bibinfo {author}
  {\bibfnamefont {H.}~\bibnamefont {Gong}}, \bibinfo {author} {\bibfnamefont
  {P.~L.}\ \bibnamefont {Venugopalan}}, \bibinfo {author} {\bibfnamefont
  {T.}~\bibnamefont {Hianik}}, \bibinfo {author} {\bibfnamefont
  {J.}~\bibnamefont {Wang}}, \ and\ \bibinfo {author} {\bibfnamefont
  {V.}~\bibnamefont {Subjakova}},\ }\bibfield  {title} {\enquote {\bibinfo
  {title} {Rapid detection of {AIB1} in breast cancer cells based on
  aptamer-functionalized nanomotors},}\ }\href@noop {} {\bibfield  {journal}
  {\bibinfo  {journal} {Chem{P}hys{C}hem}\ }\textbf {\bibinfo {volume} {20}},\
  \bibinfo {pages} {3177--3180} (\bibinfo {year} {2019})}\BibitemShut {NoStop}%
\bibitem [{\citenamefont {Barnett}\ \emph {et~al.}(2000)\citenamefont
  {Barnett}, \citenamefont {{ter Haar}}, \citenamefont {Ziskin}, \citenamefont
  {Rott}, \citenamefont {Duck},\ and\ \citenamefont {Maeda}}]{BarnettEtAl2000}%
  \BibitemOpen
  \bibfield  {author} {\bibinfo {author} {\bibfnamefont {S.~B.}\ \bibnamefont
  {Barnett}}, \bibinfo {author} {\bibfnamefont {G.~R.}\ \bibnamefont {{ter
  Haar}}}, \bibinfo {author} {\bibfnamefont {M.~C.}\ \bibnamefont {Ziskin}},
  \bibinfo {author} {\bibfnamefont {H.~D.}\ \bibnamefont {Rott}}, \bibinfo
  {author} {\bibfnamefont {F.~A.}\ \bibnamefont {Duck}}, \ and\ \bibinfo
  {author} {\bibfnamefont {K.}~\bibnamefont {Maeda}},\ }\bibfield  {title}
  {\enquote {\bibinfo {title} {International recommendations and guidelines for
  the safe use of diagnostic ultrasound in medicine},}\ }\href@noop {}
  {\bibfield  {journal} {\bibinfo  {journal} {Ultrasound in Medicine \&
  Biology}\ }\textbf {\bibinfo {volume} {26}},\ \bibinfo {pages} {355--366}
  (\bibinfo {year} {2000})}\BibitemShut {NoStop}%
\bibitem [{\citenamefont {{Holmes}}\ \emph {et~al.}(2011)\citenamefont
  {{Holmes}}, \citenamefont {{Parker}},\ and\ \citenamefont
  {{Povey}}}]{HolmesPP2011}%
  \BibitemOpen
  \bibfield  {author} {\bibinfo {author} {\bibfnamefont {M.~J.}\ \bibnamefont
  {{Holmes}}}, \bibinfo {author} {\bibfnamefont {N.~G.}\ \bibnamefont
  {{Parker}}}, \ and\ \bibinfo {author} {\bibfnamefont {M.~J.~W.}\ \bibnamefont
  {{Povey}}},\ }\bibfield  {title} {\enquote {\bibinfo {title} {Temperature
  dependence of bulk viscosity in water using acoustic spectroscopy},}\
  }\href@noop {} {\bibfield  {journal} {\bibinfo  {journal} {Journal of
  Physics: Conference Series}\ }\textbf {\bibinfo {volume} {269}},\ \bibinfo
  {pages} {012011} (\bibinfo {year} {2011})}\BibitemShut {NoStop}%
\bibitem [{\citenamefont {Weller}\ \emph {et~al.}(1998)\citenamefont {Weller},
  \citenamefont {Tabor}, \citenamefont {Jasak},\ and\ \citenamefont
  {Fureby}}]{WellerTJF1998}%
  \BibitemOpen
  \bibfield  {author} {\bibinfo {author} {\bibfnamefont {H.~G.}\ \bibnamefont
  {Weller}}, \bibinfo {author} {\bibfnamefont {G.}~\bibnamefont {Tabor}},
  \bibinfo {author} {\bibfnamefont {H.}~\bibnamefont {Jasak}}, \ and\ \bibinfo
  {author} {\bibfnamefont {C.}~\bibnamefont {Fureby}},\ }\bibfield  {title}
  {\enquote {\bibinfo {title} {A tensorial approach to computational continuum
  mechanics using object-oriented techniques},}\ }\href@noop {} {\bibfield
  {journal} {\bibinfo  {journal} {Computers in Physics}\ }\textbf {\bibinfo
  {volume} {12}},\ \bibinfo {pages} {620--631} (\bibinfo {year}
  {1998})}\BibitemShut {NoStop}%
\bibitem [{\citenamefont {Happel}\ and\ \citenamefont
  {Brenner}(1991)}]{HappelB1991}%
  \BibitemOpen
  \bibfield  {author} {\bibinfo {author} {\bibfnamefont {J.}~\bibnamefont
  {Happel}}\ and\ \bibinfo {author} {\bibfnamefont {H.}~\bibnamefont
  {Brenner}},\ }\href@noop {} {\emph {\bibinfo {title} {Low {R}eynolds Number
  Hydrodynamics: With Special Applications to Particulate Media}}},\ \bibinfo
  {edition} {2nd}\ ed.,\ \bibinfo {series} {Mechanics of Fluids and Transport
  Processes}, Vol.~\bibinfo {volume} {1}\ (\bibinfo  {publisher} {Kluwer
  Academic Publishers},\ \bibinfo {address} {Dordrecht},\ \bibinfo {year}
  {1991})\BibitemShut {NoStop}%
\bibitem [{\citenamefont {Vo{\ss}}\ and\ \citenamefont
  {Wittkowski}(2018)}]{VossW2018}%
  \BibitemOpen
  \bibfield  {author} {\bibinfo {author} {\bibfnamefont {J.}~\bibnamefont
  {Vo{\ss}}}\ and\ \bibinfo {author} {\bibfnamefont {R.}~\bibnamefont
  {Wittkowski}},\ }\bibfield  {title} {\enquote {\bibinfo {title} {Hydrodynamic
  resistance matrices of colloidal particles with various shapes},}\
  }\href@noop {} {\bibfield  {journal} {\bibinfo  {journal} {arXiv:1811.01269}\
  } (\bibinfo {year} {2018})}\BibitemShut {NoStop}%
\bibitem [{\citenamefont {Vo{\ss}}\ \emph {et~al.}(2019)\citenamefont
  {Vo{\ss}}, \citenamefont {Jeggle},\ and\ \citenamefont
  {Wittkowski}}]{VossJW2019}%
  \BibitemOpen
  \bibfield  {author} {\bibinfo {author} {\bibfnamefont {J.}~\bibnamefont
  {Vo{\ss}}}, \bibinfo {author} {\bibfnamefont {J.}~\bibnamefont {Jeggle}}, \
  and\ \bibinfo {author} {\bibfnamefont {R.}~\bibnamefont {Wittkowski}},\
  }\bibfield  {title} {\enquote {\bibinfo {title} {{HydResMat} -- {FEM}-based
  code for calculating the hydrodynamic resistance matrix of an
  arbitrarily-shaped colloidal particle},}\ }\href@noop {} {\bibfield
  {journal} {\bibinfo  {journal} {Zenodo}\ } (\bibinfo {year} {2019})},\
  \bibinfo {note} {{DOI:} 10.5281/zenodo.3541588}\BibitemShut {NoStop}%
\bibitem [{\citenamefont {Venugopalan}\ \emph {et~al.}(2020)\citenamefont
  {Venugopalan}, \citenamefont {{Esteban-Fern{\'a}ndez de {\'A}vila}},
  \citenamefont {Pal}, \citenamefont {Ghosh},\ and\ \citenamefont
  {Wang}}]{Venugopalan2020}%
  \BibitemOpen
  \bibfield  {author} {\bibinfo {author} {\bibfnamefont {P.}~\bibnamefont
  {Venugopalan}}, \bibinfo {author} {\bibfnamefont {B.}~\bibnamefont
  {{Esteban-Fern{\'a}ndez de {\'A}vila}}}, \bibinfo {author} {\bibfnamefont
  {M.}~\bibnamefont {Pal}}, \bibinfo {author} {\bibfnamefont {A.}~\bibnamefont
  {Ghosh}}, \ and\ \bibinfo {author} {\bibfnamefont {J.}~\bibnamefont {Wang}},\
  }\bibfield  {title} {\enquote {\bibinfo {title} {Fantastic voyage of
  nanomotors into the cell},}\ }\href@noop {} {\bibfield  {journal} {\bibinfo
  {journal} {ACS Nano}\ }\textbf {\bibinfo {volume} {14}},\ \bibinfo {pages}
  {9423--9439} (\bibinfo {year} {2020})}\BibitemShut {NoStop}%
\end{thebibliography}%
	
\end{document}